\documentclass[sigconf,screen]{acmart}

\AtBeginDocument{%
  \providecommand\BibTeX{{%
    \normalfont B\kern-0.5em{\scshape i\kern-0.25em b}\kern-0.8em\TeX}}}


\setcopyright{acmcopyright}
\acmPrice{15.00}
\acmDOI{10.1145/3533767.3534370}
\acmYear{2022}
\copyrightyear{2022}
\acmSubmissionID{issta22main-p47-p}
\acmISBN{978-1-4503-9379-9/22/07}
\acmConference[ISSTA '22]{Proceedings of the 31st ACM SIGSOFT International Symposium on Software Testing and Analysis}{July 18--22, 2022}{Virtual, South Korea}
\acmBooktitle{Proceedings of the 31st ACM SIGSOFT International Symposium on Software Testing and Analysis (ISSTA '22), July 18--22, 2022, Virtual, South Korea}

\usepackage{amsmath,amssymb,amsfonts}

\usepackage{hyperref}
\usepackage{algorithmic}

\usepackage{graphicx}
\usepackage{textcomp}
\usepackage{lipsum}
\usepackage{multirow}
\usepackage{multicol}
\usepackage{xcolor}
\usepackage{xspace}
\usepackage{url}
\usepackage{booktabs}
\usepackage{tcolorbox}
\usepackage{array}
\usepackage{tabularx}
\usepackage{makecell}
\usepackage{listings}
\usepackage[ruled,vlined,linesnumbered]{algorithm2e}

\definecolor{pblue}{rgb}{0.13,0.13,1}
\definecolor{pgreen}{rgb}{0,0.5,0}
\definecolor{pred}{rgb}{0.9,0,0}
\definecolor{pgrey}{rgb}{0.46,0.45,0.48}
\definecolor{light-gray}{gray}{0.85}
\newcommand{\metric}[1]{\text{#1}}

\def\BibTeX{{\rm B\kern-.05em{\sc i\kern-.025em b}\kern-.08em
    T\kern-.1667em\lower.7ex\hbox{E}\kern-.125emX}}
\def\splitmetric{\metric{Split}\xspace}
\def\covermetric{\metric{Cover}\xspace}
\def\unifiedmetric{\metric{FDG}\xspace}
\def\dfj{\textsc{Defects4J}\xspace}
\def\evo{\textsc{EvoSuite}\xspace}
\def\Tnt{T \cup \left\{t\right\}}
\def\centerbullet{\multicolumn{1}{c|}{$\bullet$}}

\DeclareMathOperator*{\argmax}{argmax}

\begin{document}

\title{FDG: A Precise Measurement of Fault Diagnosability Gain of Test Cases}

\author{Gabin An}
\affiliation{%
  \institution{School of Computing, KAIST}
  \city{Daejeon}
  \country{Republic of Korea}
}
\email{agb94@kaist.ac.kr}

\author{Shin Yoo}
\affiliation{%
  \institution{School of Computing, KAIST}
  \city{Daejeon}
  \country{Republic of Korea}
}
\email{shin.yoo@kaist.ac.kr}

\begin{abstract}
The performance of many Fault Localisation (FL) techniques directly depends on the 
quality of the used test suites. Consequently, it is extremely useful to be 
able to precisely measure how much diagnostic power each test case can 
introduce when added to a test suite used for FL. Such a measure can help 
us not only to prioritise and select test cases to be used for FL, 
but also to effectively augment test suites that are too weak to be used with 
FL techniques. We propose \unifiedmetric, a new measure of Fault
Diagnosability Gain for individual test cases. The design of \unifiedmetric is 
based on our analysis of existing metrics that are designed to 
prioritise test cases for better FL. Unlike other metrics, 
\unifiedmetric exploits the ongoing FL results to emphasise 
the parts of the program for which more information is needed.
Our evaluation of 
\unifiedmetric with \dfj shows that it can successfully help the augmentation of
test suites for better FL.
When given only a few failing test cases (2.3 test cases on 
average), \unifiedmetric can effectively augment the given test suite by 
prioritising the test cases generated automatically by \evo: the augmentation 
can improve the acc@1 and acc@10 of the FL results by 11.6x and 2.2x on 
average, after requiring only ten human judgements on the correctness of the 
assertions \evo generates.
\end{abstract}

\maketitle

\section{Introduction}
\label{sec:introduction}

Fault Localisation (FL) techniques aim to reduce the cost of software debugging 
by automatically identifying the root cause of the observed test failures~\cite{
wong2016survey}. While some FL techniques use only static data such as lexical 
proximity between bug reports and source code~\cite{Saha:2013rm}, the majority 
of FL techniques are based on dynamic analysis: Spectrum Based Fault 
Localisation (SBFL) relies on both the test coverage and the test results~\cite{
jones2005empirical, naish2011model,wong2014tor}, whereas Mutation Based Fault 
Localisation exploits the relationship between mutants and test cases~\cite{
Moon:2014ly,Hong:2017qy,Papadakis:2015sf,Papadakis:2012fk}. In both cases, the 
quality of the test suite used with an FL technique can significantly affect 
its performance. It is widely known that FL techniques thrive only when 
accompanied by a rich and diverse test suite~\cite{perez2017test}.

In reality, however, it is not always the case that the given test suite is 
sufficiently diverse to ensure successful fault localisation.
In fact, it is 
not unusual to start the debugging process only with a single failing test 
input, calling for the need of test augmentation. Many techniques have been 
developed to guide the test augmentation, i.e., to add the test case that can 
maximise the diagnosability of the augmented test suite~\cite{campos2013entropy,
artzi2010directed}. Equally relevant are the techniques that prioritise the 
given test cases for better and earlier fault localisation~\cite{yoo2013fault,
gonzalez2010prioritizing}, as well as techniques that aim to measure the 
diagnostic capability of the given test suite. Most of these techniques are 
built around a metric that measures the diagnosability of either a single test, 
or a set of tests. However, our analysis of existing test diagnosability 
metrics reveals room for improvement. Specifically, while some of the 
techniques are \emph{result-aware}, i.e., incorporate the result of test 
executions (pass/fail) into the diagnosability computation, none of them 
directly uses the suspiciousness scores of individual program elements 
\emph{during localisation}, even though they can provide critical information 
by pinpointing where the prioritisation should focus on next.

Based on our analysis of existing techniques, we propose \unifiedmetric, a 
metric that can precisely measure the \textbf{\underline{F}}ault 
\textbf{\underline{D}}iagnosability \textbf{\underline{G}}ain that a test case 
can bring to a test suite. \unifiedmetric is designed to be used \emph{during} 
fault localisation, and uses the current suspiciousness scores to precisely 
capture which part of the target program requires \textit{additional diagnostic 
information} from the test case under consideration. We first evaluate 
\unifiedmetric with developer-written test cases in \dfj, i.e., under existing 
and reliable test oracles and with all test coverage available, to study its performance under an ideal setting. We construct fixed-sized test suites by adding test 
cases in the descending order of their diagnosability measured by \unifiedmetric and other existing metrics. 
Test suites constructed based on \unifiedmetric 
produce much more accurate fault localisation, achieving acc@10 values that are 
23\% and 14\% higher than those produced by the state-of-the-art metrics.

In a more realistic setting, we also evaluate \unifiedmetric using 
automatically generated test cases to augment the initial set of failing test 
inputs. We posit that the true cost of test suite augmentation is not the cost of test data generation, as it can be automated. Rather, the major cost of test 
augmentation is the human effort required to produce the test oracles for the 
generated test data. Our evaluation of \unifiedmetric assumes an iterative FL scenario, in which an 
insufficient test suite is augmented, on the fly, with test data automatically 
generated by \evo. Here, \unifiedmetric is used to choose the next test case to 
be presented to the human engineer for the oracle labelling. We show that after 
only ten interactions with the human engineer, \unifiedmetric can boost acc@1 
and acc@10 by 11.6 (from 5 to 58) and 2.2 (from 63 to 147) times, respectively, 
compared to the initial test suites that contain only a few failing test cases.
In addition, we also show that \unifiedmetric is fairly resilient to labelling 
errors with an error rate of up to 30\%.

The main contributions of this paper are as follows:
\begin{itemize}
\item We analyse, and empirically compare, existing metrics that measure 
the diagnosability of test cases for fault localisation. Our evaluation 
compares how quickly existing metrics can improve the accuracy of SBFL 
techniques by prioritising the human-written test cases in \dfj.

\item We propose a novel diagnosability metric for fault localisation,
\unifiedmetric, based on the analysis of existing metrics. \unifiedmetric 
distinguishes itself from existing metrics in that it actively uses the 
available suspiciousness scores to capture which part of the target
program needs additional diagnostic information while localisation is ongoing. 

\item We first evaluate \unifiedmetric with human-written tests, to 
investigate its performance under perfect test oracles. When forced to use a 
limited number of test cases, using those chosen by \unifiedmetric can achieve 
up to 23\% higher acc@10 when compared to the localisation driven by 
state-of-the-art metrics.

\item We also report findings from a more realistic scenario, where the 
developer labels oracles that are generated by \evo to augment
an initial set of failing inputs
for fault localisation: we use \unifiedmetric to choose which test data to 
present to the human engineer for test oracle labelling.
\unifiedmetric can boost acc@1 by 11.6 times after only ten human labels.

\item We make our replication package publicly available at:
\url{https://github.com/agb94/FDG-artifact}~\cite{fdg-artifact}
\end{itemize}

The remainder of the paper is organised as follows. Section~\ref{sec:background}
presents the basic notations that will be used throughout the paper, as well 
as the fundamental concepts in SBFL. Section~\ref{sec:method_study} analyses
existing diagnosability metrics and presents a comparison of their performance 
using human-written tests in \dfj. Section~\ref{sec:ours} proposes our novel 
metric, \unifiedmetric, based on our analysis. Section~\ref{sec:exp_setup} 
describes the experimental setup for our empirical evaluation, the results of 
which are presented in Section~\ref{sec:results}. Section~\ref{sec:discussion} discusses implications of this work, while Section~\ref{sec:relatedwork} 
presents the related work, and Section~\ref{sec:threats} considers threats to 
validity. Finally, Section~\ref {sec:conclusion} concludes.

\section{Preliminaries}
\label{sec:background}

This section introduces the basic notations, the background of SBFL techniques,
and the definition of ambiguity groups.

\subsection{Basic Notation}

Given a program $P$, let us define the following:

\begin{itemize}
\item Let $E = \{e_1, \ldots, e_n\}$ be the set of program elements that consist $P$, such as
statements or methods, and $T = \{t_1, \ldots, t_m\}$ be the test suite for $P$.
\item Let $C_T$ be the $m \times n$ coverage matrix of $T$:
$$
C_T=\left[\begin{array}{cccc}
a_{11} & \cdots & a_{1n} \\
%c_{2,1} & c_{2,2} & \cdots & c_{2,n} \\
\vdots & \ddots & \vdots \\
a_{m1} & \cdots & a_{mn}
\end{array}\right] \in \mathbb{B}^{m \times n}
$$
where $C_T[i, j] = a_{ij} = 1$ if $t_i$ covers $e_j$ and 0 otherwise. 
\item Given an arbitrary test case $t$, let $c_t \in \mathbb{B}^{1 \times n}$
be the coverage vector of $t$, where $c_t[j]= 1$ if $t$ executes $e_j$, and 0 otherwise.

\item Let $R: T \rightarrow \mathbb{B}$ be the function that maps a test in
$T$ to its result: $R(t) = 0$ if $t$ reveals a fault in $P$, and 1 otherwise.
Subsequently, the set of failing tests, $T_f$, can be defined as $\left\{t \in T|R(t) = 0\right\}$.
$P$ is \textit{faulty} when $T_f$ is not empty.

\end{itemize}

\subsection{Spectrum-based Fault Localisation (SBFL)}
SBFL is a statistical approach for
the automated localisation of software faults~\cite{wong2016survey}. It utilises
\textit{program spectrum}, a summary of the runtime information collected
from the program executions, to find the faulty program elements.
A representative program spectrum of a program element $e_j \in E$ consists of four values:
$(e_p, n_p, e_f, n_f) = (N_{11}, N_{01}, N_{10}, N_{00})$
where 
$$
N_{ab} = |\left\{t_i \in T | C_T[i,j] = a \land R(t_i) = b \right\}|
$$

SBFL statistically estimates the suspiciousness of each program element based 
on the following rationale: \textit{``The more the execution of an element is 
correlated with failing tests (high $e_f$, low $n_f$) and less with passing 
ones (high $n_p$, low $e_p$), the more suspicious the program element is, and 
vice versa''}.
A risk evaluation formula $S:\mathbb{I}^4\rightarrow\mathbb{R}$ is a formula 
that converts the four program spectrum values to a suspiciousness score. Many risk 
evaluation 
formulas~\cite{abreu2006evaluation, jones2005empirical, naish2011model, 
wong2014tor} have been designed to implement this core idea. For 
example, Ochiai~\cite{abreu2006evaluation}, one of the most widely studied 
risk evaluation formulas, computes the suspiciousness
as follows: 
\begin{equation}
\label{eq:Ochiai}
Ochiai(e_p, n_p, e_f, n_f) = e_f/\sqrt{(e_f+n_f)\times(e_f+e_p)}
\end{equation}
Fig.~\ref{fig:moti_ex} illustrates a concrete example of calculating Ochiai
scores from the coverage matrix and test results. 

SBFL is one of the most widely studied FL techniques~\cite{wong2016survey} as 
it is applicable as long as test coverage is available. Due to the 
applicability, SBFL scores are often used as features for more complicated 
learn-to-rank FL techniques~\cite{Sohn:2017xq,Li2019aa,B.-Le:2016yu}, 
motivating us to improve its effectiveness.

\subsection{Ambiguity Groups}
An \textit{ambiguity group}~\cite{stenbakken1989ambiguity, gonzalez2011prioritizing}
is a set of program elements that are only executed by the same set of test cases.
Given a program $P$ and its elements $E$, we define $AG(T)$ as a set of such
ambiguity groups under the test suite $T$. The size of $AG(T)$ is equal
to the number of unique columns in $C_T$. More formally, $AG(T)$ is a partition~\cite{halmos2017naive} of the set $E$ that satisfies the following properties:

\begin{enumerate}
\item $\forall e \in E \ldotp \exists g \in AG(T) \ldotp e \in g$
\item $\forall g \in AG(T) \ldotp \forall e_i, e_j \in g \ldotp C_T[:, i] = C_T[:, j]$
\item $\forall g_1, g_2 \in AG(T) \ldotp g_1 \ne g_2 \land e_i \in g_1 \land e_j \in g_2 \\\implies C_T[:, i] \ne C_T[:, j]$
\end{enumerate}

Elements in the same ambiguity groups are assigned the same suspiciousness 
score, since their program spectra are identical. For example, in 
Fig.~\ref{fig:moti_ex}, $AG(\{t_1, t_2, t_3\})$ is $\left\{\left\{e_1, e_2\right\}, \left\{e_3, e_4\right\}, \left\{e_5\right\}\right\}$, and all the
elements in an ambiguity group are tied in the final ranking and hence cannot be 
uniquely diagnosed as faulty. Therefore, the performance of an SBFL technique 
increases when there are more ambiguity groups, and their size is smaller.

\renewcommand{\arraystretch}{0.85}
\begin{figure}[t]
\centering
\scalebox{0.85}{
\begin{tabular}{l|c|c|c|c|c|c|c}
\toprule
\multicolumn{2}{l|}{Program Elements}  & $e_1$ & $e_2$ & {\color{red}$e_3$} & $e_4$ & $e_5$ & \\
\midrule
\multirow{3}{*}{Tests}                             & $t_1$  &           &           &           &           & \centerbullet & PASS              \\
                                                   & $t_2$  & \centerbullet & \centerbullet & \centerbullet & \centerbullet &           & FAIL               \\
                                                   & $t_3$  &           &  & \centerbullet & \centerbullet &           & FAIL               \\ \midrule
\multirow{4}{*}{Spectrum}                          & $e_p$  & 0         & 0         & 0         & 0         & 1         &                     \\
                                                   & $n_p$  & 1         & 1         & 1         & 1         & 0         &                     \\
                                                   & $e_f$  & 1         & 1         & 2         & 2         & 0         &                     \\
                                                   & $n_f$  & 1         & 1         & 0         & 0         & 2         &                     \\ \midrule
\multicolumn{2}{l|}{Ochiai}                         & 0.71   & 0.71      & 1.00       & 1.00      & 0.00      &           \\ \midrule
\multicolumn{2}{l|}{Rank}                           & 4      & 4         & 2         & 2         & 5         &           \\ \midrule\midrule
\multirow{4}{*}{\shortstack[l]{Additional\\Tests}} & $t_1'$ &           &           &           &           & \centerbullet & PASS                  \\
                                                   & $t_2'$ & \centerbullet &           &           &  & \centerbullet &  PASS          \\
                                                   & $t_3'$ & \centerbullet & \centerbullet & \centerbullet & \centerbullet &           & FAIL   \\
                                                   & $t_4'$ &           &           & \centerbullet &           &           & FAIL           \\\midrule
\multirow{4}{*}{Ochiai}& w/ $t_1'$ & 0.71   & 0.71      & 1.00       & 1.00      & 0.00      &           \\
                       & w/ $t_2'$ & 0.50   & 0.71      & 1.00       & 1.00      & 0.00      &           \\
                       & w/ $t_3'$ & 0.82   & 0.82      & 1.00       & 1.00      & 0.00      &           \\
                       & w/ $t_4'$ & 0.71   & 0.71      & \textbf{1.00}       & 0.82      & 0.00      &           \\\bottomrule
\end{tabular}}
\caption{A simple motivating example with one faulty element $e_3$. The dots ($\bullet$) show the coverage relation, and the ranks are computed using the max tie-breaker.\label{fig:moti_ex}}
\Description{A simple motivating example (The dots ($\bullet$) show the coverage relation, and the ranks are computed using the max tie-breaker.)}
\end{figure}
\renewcommand{\arraystretch}{1.5}

\section{Analysis of Existing Diagnosibility Metrics}
\label{sec:method_study}

Fig.~\ref{fig:moti_ex} contains a motivating example that shows the role of 
a diagnosability metric in FL. In the existing SBFL results based on 
original test cases $\{t_1, t_2, t_3\}$, program elements $e_3$ and $e_4$ 
form an ambiguity group. They share the highest suspiciousness score, i.e., 
1.0, even though only $e_3$ is the actual faulty element.
In this case, to further distinguish which one is more suspicious between 
them, we need an additional 
test case that can break the ambiguity group, i.e., covering either one of
the program elements.

Let us now consider the four additional test cases, $t_1'$ to $t_4'$, to 
augment the existing test suite. Adding $t_1'$ or 
$t_3'$ does not provide additional diagnostic information, as the program elements in the same ambiguity group are 
either executed or not executed together. In comparison, $t_2'$ and $t_4'$ can 
provide more information because they can break an ambiguity group,
$\{e_1, e_2\}$ and $\{e_3, e_4\}$, respectively. Between them, $t_4'$ should be 
prioritised over $t_2'$ since the ambiguity group $\{e_3, e_4\}$ is currently more suspicious than
$\{e_1, e_2\}$. Fig.~\ref{fig:moti_ex} shows how Ochiai scores change
as each test case is added to the original test suite. When $t_4'$ is added,
the faulty element $e_3$ is finally ranked at top as the suspiciousness score 
of the non-faulty element $e_4$ decreases.
Quantifying such diagnostic contribution made by individual test cases can 
inform our test suite augmentation, or even the order of test execution.

There are many existing diagnosability metrics that are designed to 
capture the information gain contributed by individual test cases. In many 
cases, these metrics are used as part of a test prioritisation technique that 
is designed to achieve more accurate fault localisation earlier, or with fewer 
test cases~\cite{hao2010test,Gonzalez-Sanchez:2011yq,yoo2013fault}. We can 
also derive such metrics from diagnosability measures of the entire test 
suites~\cite{perez2017test,baudry2006improving} by looking at the difference in 
diagnosability gain achieved by the addition a single test. To the best of 
our knowledge, these metrics have never been directly compared to each other 
using the same benchmark. The remainder of this section describes 
existing diagnosability metrics, and empirically compares them using 
\dfj, a collection of real-world faults in Java programs.

\subsection{Analysis of Existing Diagnosability Metrics}
\label{sec:survey}

We analyse nine diagnosability metrics for SBFL from the 
relevant literature published after 2010. Two of these are simply coverage 
based Test Case Prioritisation~\cite{yoo2012regression} techniques that serve 
as baselines. Out of the seven remaining metrics, four metrics have been 
introduced as part of test prioritisation techniques that aim to improve FL 
performance, while the remaining three are derived from the diagnostic 
capability metrics for entire test suites. Given a diagnosability metric
$f$, we use the notation $f(T, t)$ to denote the estimated diagnosability gain 
brought by a newly chosen single test case $t$ to the set of already executed 
test cases $T$.

\subsubsection{Test Case Prioritisation Metrics}
\label{sec:prioritisation_metrics}

Since SBFL techniques are based on an aggregation of test coverage, we 
include two coverage-based test case prioritisation techniques, total and 
additional~\cite{rothermel1999test, elbaum2002test}, as baselines. Total 
prioritisation greedily selects the test case with the highest coverage over 
all program elements, while additional prioritisation favours the test case 
with the highest coverage over remaining uncovered program elements only: 
$$
%\metric{Total}(T, t) = |\left\{1 \leq j \leq n|c_t[j] = 1\right\}|
\metric{Total}(T, t) = |\left\{e_j \in E|c_t[j] = 1\right\}|
$$
$$
\metric{Add}(T, t) = |\left\{e_j \in E|c_t[j] = 1 \land \forall t_i \in T \ldotp C_{T}[i, j] = 0\right\}|
$$
In both cases, our rationale is that increasing coverage as early as possible may 
lead to an earlier increase in the amount of information provided to SBFL 
techniques.

Renieris and Reiss proposed Nearest Neighbours fault localisation~\cite{renieres2003fault},
which selects the passing test cases whose
execution trace is the closest to the given failing execution. They use a 
program-dependency-based metric to measure the proximity. Motivated by this 
work, Bandyopadhyay et al. put weights to test cases using the average of 
Jaccard similarity between the coverage of a given test case and the failing 
ones~\cite{bandyopadhyay2011proximity}. Although Bandyopadhyay et al. do not 
use this metric to prioritise test cases, we include the metric since it
aims to measure the diagnosability of individual test cases. We call this metric
as Prox hereafter:
$$
\metric{Prox}(T, t) = \frac{\sum_{t_i \in T_f}Jaccard(C_T[i,:], c_t)}{|T_f|}
%\metric{Prox}(T, t) = \sum_{t_i \in T_f}Jaccard(C_T[i,:], c_t)/|T_f|
$$

Hao et al.~\cite{hao2010test} select a representative subset 
of test cases from a given test suite to reduce the oracle cost (i.e., to 
reduce the number of outputs to inspect). Among the three coverage-based 
strategies introduced, S1, S2, and S3, the most effective one for fault
localisation was S3: it takes 
into account the relative importance of ambiguity groups, which in turn is 
defined as the ratio of failing tests that execute each group, $pri(g) = e_f/(e_f + n_f)$.
$$
\metric{S3}(T, t) = \sum_{g \in AG'(T)} pri(g) \cdot div(t, g)
$$
where $AG'$ denotes a variation of ambiguity group, which includes 
program elements that share the same spectrum values, and 
$div(t, g)$ denotes the size of the smaller subset when $t$ 
is used to divide $g$ to two subsets based on coverage, i.e.,
the minimum between 
$|\{e_j \in g | c_t[j] = 1\}|$ and $|\{e_j \in g | c_t[j] = 0\}|$. 
Therefore, S3 assigns higher priority to tests that can more evenly split the 
more important ambiguity groups.

RAPTER~\cite{gonzalez2011prioritizing} prioritises test cases by the amount of 
ambiguity group reduction achieved by individual test cases. While the aim of 
reducing ambiguity group is similar to S3, RAPTER considers only the size of 
ambiguity groups, and not the test results.\footnote{S3, on the other hand,
considers test results via $pri(g)$.} Here, $p(g) = |g|/n$ is the 
probability that $g$ contains a faulty element, which is assigned in proportion to the size of the group, and $\frac{|g|-1}{2}$ refers 
to the expected wasted effort if a developer considers program elements in $g$ 
randomly.
$$
\metric{RAPTER}(T, t) = -\sum_{g \in AG(\Tnt)}p(g)\cdot\frac{|g|-1}{2}
$$

FLINT~\cite{yoo2013fault} is an information-theoretic approach that formulates
test case prioritisation as an entropy reduction process. In FLINT,
a test case is given higher priority when it is expected to reduce more 
entropy (H)~\cite{shannon1948mathematical} in the suspiciousness distribution 
across program elements. The expected entropy reduction of each test case is
predicted based on the conditional probability of the test case failing; the 
probability of a new test case failing is approximated as the failure rate observed among the test cases executed so far.
$$
\metric{FLINT}(T, t) = - \alpha \cdot H(P_f) - (1 - \alpha) \cdot H(P_p)
$$

Here, $\alpha$ is the observed failure rate, $|T_f|/|T|$, and $P_p$ and $P_f$ 
are the suspiciousness distributions of $\Tnt$, computed under the assumption 
that $t$ passes and fails, respectively. The suspiciousness distributions are 
computed by normalising the Tarantula~\cite{jones2005empirical} scores. Among 
the studied existing metrics, FLINT is the only one that does use 
suspiciousness scores to measure the diagnosability of a new test to execute. 
However, it only considers the overall distribution of suspiciousness via 
Shannon entropy, and does not use scores of individual program elements to 
focus the prioritisation.
Note that we negate the original RAPTER and FLINT metrics so that a higher score means a higher diagnosability gain.

\subsubsection{Test Suite Diagnosability Metrics}
\label{sec:diagnosability_metrics}

This section describes test suite diagnosability metrics that are designed to 
measure the diagnosability of entire test suites. Despite originally being designed for test suites, a test suite 
diagnosability metric $F$ can also be used to quantify the diagnosability 
\textit{gain} of an individual test, $t$, by computing the difference between 
the diagnosability of an original test suite, $F(T)$, and that of the enhanced 
test suite, $F(\Tnt)$:
$$f(T, t) = F(\Tnt) - F(T)$$

Baudry et al.~\cite{baudry2006improving} analysed the features of a test suite 
that are related to the fault diagnosis accuracy and introduced the \textit{
Test-for-Diagnosis (TfD)} metric that measures the number of ambiguity groups
(referred to as \textit{Dynamic Basic Blocks (DBB)} in their paper).
They proposed composing a test suite that maximises the number of DBBs
for high diagnosability.
$$
\metric{TfD}(T) = |AG(T)|
$$

EntBug~\cite{campos2013entropy} evaluates a test suite based on its coverage matrix 
density, which is defined as the ratio of ones in the coverage matrix.
EntBug augments an existing test suite with additionally 
generated test cases with the goal of balancing the density of the coverage 
matrix to 0.5.
$$
\metric{EntBug}(T) = 1 - |1 - 2 \cdot \rho \left(T\right)|
$$

where $\rho\left(T\right) = \sum C_T[i, j]/(|E|\cdot|T|)$. Note that this 
definition is the normalised version~\cite{perez2017test} of EntBug.

More recently, Perez et al. propose DDU~\cite{perez2017test}, a test suite 
diagnosability metric for SBFL, that combines three key properties, \textbf{d}ensity, \textbf{d}iversity, and \textbf{u}niqueness, all being properties that 
a test suite should exhibit to achieve high localisation accuracy. The density 
component is identical to EntBug, while the uniqueness component is the ratio 
of the number of ambiguity groups over the number of all program elements, i.e., $
|AG(T)|/|E|$. Lastly, the diversity component is designed to ensure the 
diversity of test executions, i.e., the contents of the rows in the coverage 
matrix. Formally, it is defined as the Gini-Simpson index~\cite{jost2006entropy} among the rows.
$$
\metric{DDU}(T) = \metric{density}(T) \times \metric{diversity}(T) \times \metric{uniqueness}(T)
$$

\subsubsection{Classification of Diagnostic Capability Metrics}
\label{sec:metric_classification}

Among the aforementioned metrics, some can be calculated with only the 
test coverage, while others require the test results plus the coverage.
Thereby, we broadly 
classify them into two categories based on the utilised information:

\begin{itemize}
\item \textbf{Result-Agnostic}: metrics utilising only coverage information; total
coverage, additional coverage, RAPTER, TfD, EntBug, and DDU belong to this category.
\item \textbf{Result-Aware}: metrics utilising coverage information and previous test results; Prox, S3, and FLINT belong here.
\end{itemize}

When we order tests according to diagnosability metrics, the result-agnostic
metrics are not affected by the results of previously chosen, whereas the result-aware metrics are. 

\subsection{Evaluation of Existing Metrics}

We empirically compare and analyse the performance of existing metrics by 
studying how much each metric can accelerate fault localisation of real world 
bugs. For this, we apply the studied existing metrics to the bugs and 
human-written test cases in the version 2.0 of \dfj, a widely studied real 
world fault benchmark (for details about subject programs and bugs, please 
refer to Section~\ref{sec:subjects}). We adopt a ranking-based evaluation 
protocol stated below: while there are questions about whether the ranking form 
is the most effective way for humans to consume results of FL 
techniques~\cite{parnin2011automated}, we posit that test suites capable of 
producing more accurate rankings are also capable of providing higher 
diagnosibility due to the diversity in its coverage.

\subsubsection{Protocol}
\label{sec:comparison_protocol}

For every studied fault, we create a test suite that initially only contains 
\emph{one} of the failing test cases provided by \dfj: therefore, we create 
multiple such test suites based on a single \dfj fault, if it has multiple 
failing test cases. Subsequently, we iteratively add ten test cases from the 
remaining test cases, in the order given by one of the studied existing 
metrics: after each iteration, we evaluate the accuracy of fault localisation 
performed using test cases added up to that point. In total, we study 810 test 
orderings generated from 351 studied faults.

To evaluate the SBFL accuracy at each iteration, we compute the line level 
Ochiai (Eq.~\ref{eq:Ochiai}) scores, and aggregate them at the method 
level~\cite{Sohn:2017xq} by assigning each method the score of its most 
suspicious line. We use max-tiebreaker to break ties in the ranking. Finally, 
we use Mean Average Precision (mAP) to compare rankings produced by different 
metrics. mAP is a widely used metric in Information Retrieval, and is defined 
as the Average Precision (AP) values across multiple queries (in our case, 
multiple test orderings). Average Precision, in turn, stands for the average of 
precision values at each rank of all positive samples (in our case, faulty 
methods). For example, if a program contains two faulty methods, A and B, 
placed at the second and the fifth places, the AP is $\frac{1}{2}(
\frac{1}{2} + \frac{2}{5}) = 0.45$: the higher the mAP value is, the better the 
overall ranking is.

\subsubsection{Comparison Results}
\label{sec:comparison_results}

\begin{figure}[t]
  \centerline{\includegraphics[width=\linewidth]{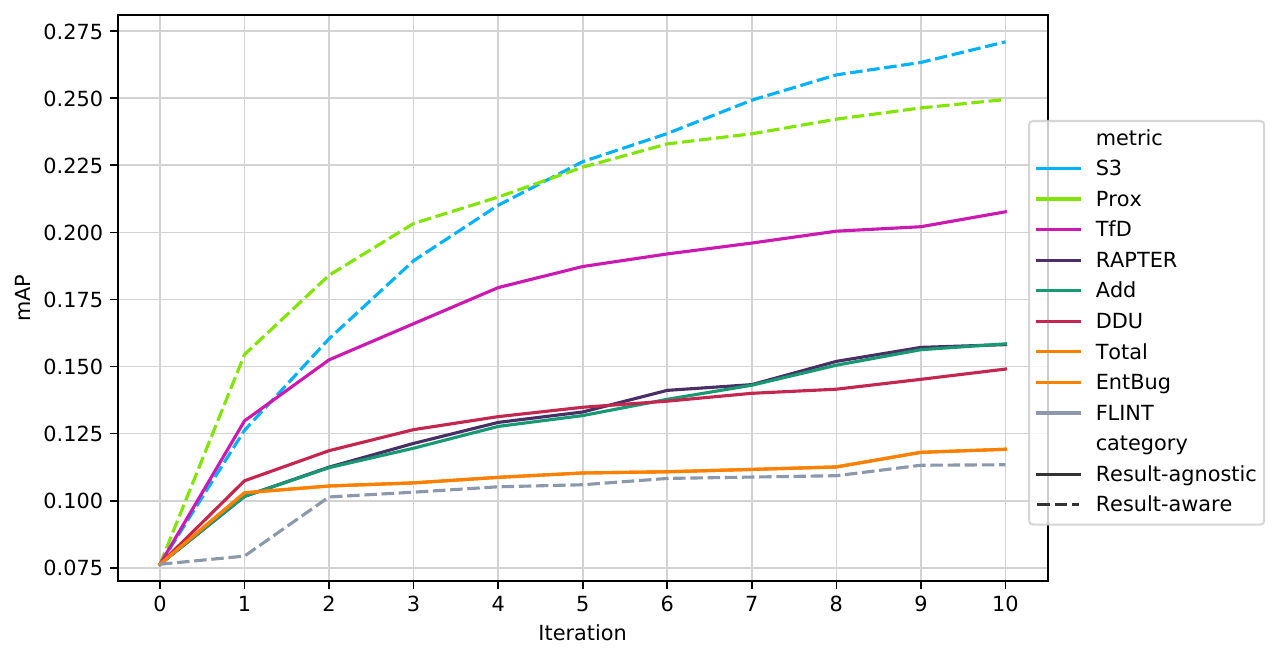}}
  \Description{The mAP values for each prioritisation metric during ten iterations}
  \caption{The mAP values for each prioritisation metric during ten iterations}
  \label{fig:existing_mAP}
\end{figure}

Figure~\ref{fig:existing_mAP} shows the trends of mAP values produced by 
different metrics at each iteration. Lines are colour-coded for each metric; 
metrics in result-agnostic and result-aware categories are shown in solid 
and dashed lines, respectively. While we only show the results for the first 
ten iterations, note that the mAP values will eventually converge to the same 
value per a \dfj fault, as the Ochiai ranking produced using all test cases 
should be identical w.r.t. a deterministic tie-breaker. A metric is more 
effective if the line converges faster. 

The results show that two result-aware metrics, S3 and Prox, outperform 
all other result-agnostic metrics. With S3, the initial mAP value, $0.076$, is 
increased to $0.271$ (an increase of 257\%) after the ten test cases are 
selected, whereas TfD, the best performing result-agnostic metric, only 
achieves 172\% improvement during the same number of iterations. Prox, which 
sorts test cases by their similarity to failing tests, shows faster initial 
convergence, while S3 achieves higher localisation performance after the fifth 
iteration. Both S3 and TfD, metrics that aim to split ambiguity groups, show 
the best result at the tenth iteration among the result-aware and 
result-agnostic metrics, respectively. Compared to S3 and Prox, FLINT does not 
perform as well, despite being a result-aware metric. FLINT is originally 
designed to initially prioritise for coverage and switch to prioritisation for 
fault localisation once a test fails, whereas our evaluation scenario starts 
with a failing test case. Consequently, the observed failure rate starts at 1.0 
for FLINT, which significantly skews its following analysis.

Among the result-agnostic metrics, Total and EntBug perform identically on our 
subjects. Total gives higher priority to tests that cover more program 
elements. Similarly, EntBug also takes into account the number of covered 
program elements, while aiming to reach the optimal density of 0.5. However, 
since the level of coverage achieved by individual human-written test cases in 
our subject is fairly low, EntBug ends up trying to increase the coverage for 
all ten iterations, which in turn increases the density. Another interesting 
observation is that TfD outperforms DDU, even though TfD is conceptually 
identical to the uniqueness component of DDU. DDU gives high priority to test 
cases that are either not similar to existing failing test cases or cover more 
elements, thanks to the diversity and the density metrics. However, the results 
of Prox show that choosing tests similar to failing tests can be effective, 
while Total and EntBug show that relying on coverage density alone may not be 
effective. Based on this, we posit that, when a failing test already exists and 
is known, it is better to prioritise tests that are similar to the failing test 
than to focus on test diversity.

In summary, our evaluation using \dfj shows that two result-aware metrics, S3 and Prox, outperform all result-agnostic metrics. When diagnosing faults that have already been observed by failing executions, result-agnostic metrics cannot efficiently prioritise test cases because they cannot distinguish the more suspicious elements from those that are less so.

\section{Fault Diagnosibility Gain}
\label{sec:ours}

This section proposes a novel diagnosability metric, \unifiedmetric
(Fault Diagnosibility Gain), that better measures the fault diagnosability gain 
by leveraging the ongoing FL results during the prioritisation.

\subsection{Design of \unifiedmetric (\underline{F}ault \underline{D}iagnosability \underline{G}ain)}

Our new diagnosability metric, \unifiedmetric, consists of two 
subcomponents, \splitmetric and \covermetric: \splitmetric is designed to break 
more suspicious ambiguity groups, and \covermetric is designed to focus on 
covering more suspicious elements. 

\subsubsection{Breaking the suspicious ambiguity groups}
\label{sec:split_metric}

\splitmetric measures the expected wasted localisation effort 
when a new test $t$ is added to $T$. It extends RAPTER by weighting 
each ambiguity group using the previous test results as in S3, instead of the 
size of the ambiguity group. However, while S3 only considers the number 
of failing test cases, \splitmetric uses suspiciousness scores, which take 
into account both the number of failing test cases and passing ones.

Let us first define the probability of an ambiguity group containing the 
fault, $p(g)$, as the sum of the probabilities of its elements being faulty:
\[ p(g) = \sum_{e_j \in g} p(e_j) \]
where $p(e_j)$ refers to the probability of each program element $e_j$ being faulty.
We obtain $p(e_j)$ by applying the \textit{softmax} function on the suspiciousness scores calculated using $T$, i.e.,
$p(e_j) = w_j/\sum_{i}w_i$, where $w_j$ is the suspiciousness score\footnote{Note that we use min-max scaled Ochiai scores throughout this paper.} of $e_j$.
Using $p(g)$, \splitmetric is defined as follows:
\begin{equation*}
\begin{split}
\splitmetric(T, t) &= 1 - \sum_{g \in AG(\Tnt)}p(g) \cdot \left(\frac{|g|-1}{2}\right) / \left(\frac{n - 1}{2}\right) \\
  &= 1 - \frac{1}{n - 1}\cdot\sum_{g \in AG(\Tnt)}p(g) \cdot (|g|-1) \\
\end{split}
\end{equation*}
Here, $(n - 1)/2$ is the normalisation constant, which is the maximum 
localisation effort when all program elements belong to a single ambiguity 
group. In general, if $\Tnt$ produces larger and more suspicious ambiguity groups, \splitmetric will penalise $t$ more heavily.

\subsubsection{Covering the suspicious program elements}
\label{sec:cover_metric}

\covermetric quantifies how much a new test case covers current suspicious 
elements. More formally, it measures weighted coverage, where the weights of program elements are simply their suspiciousness scores.

$$
\covermetric(T, t) = \frac{\sum_{j=1}^{n}w_j \cdot c_t[j]}{n}
$$

\covermetric shares motivation with Prox~\cite{renieres2003fault, bandyopadhyay2011proximity},
which directly uses coverage similarity to failing tests.
However, this direct comparison is its downfall: if there are multiple failing tests 
with significantly different coverage patterns, averaging 
similarities may not be the best way to measure the similarity between test 
cases. We instead use the suspiciousness score, which can be thought of as an
aggregation of failing and passing test executions, to weight coverage.

\subsubsection{A Combined Metric, \unifiedmetric}
\label{sec:combining_metrics}

As both \splitmetric and \covermetric are normalised, we define \unifiedmetric as the weighted sum of both\footnote{Our internal evaluation showed that adding two metrics performs better than multiplying them for aggregation.}:

\begin{equation}
\label{eq:fdg}
\unifiedmetric(T, t) = \alpha \cdot \splitmetric(T, t) + (1-\alpha)\cdot\covermetric(T, t)
\end{equation}

The coefficient $\alpha$ can be tuned using the information of known faults: we study the impact of $\alpha$ with RQ4 (see Section~\ref{sec:rqs}).

\begin{figure}[t]
  \centerline{\includegraphics[width=0.95\linewidth]{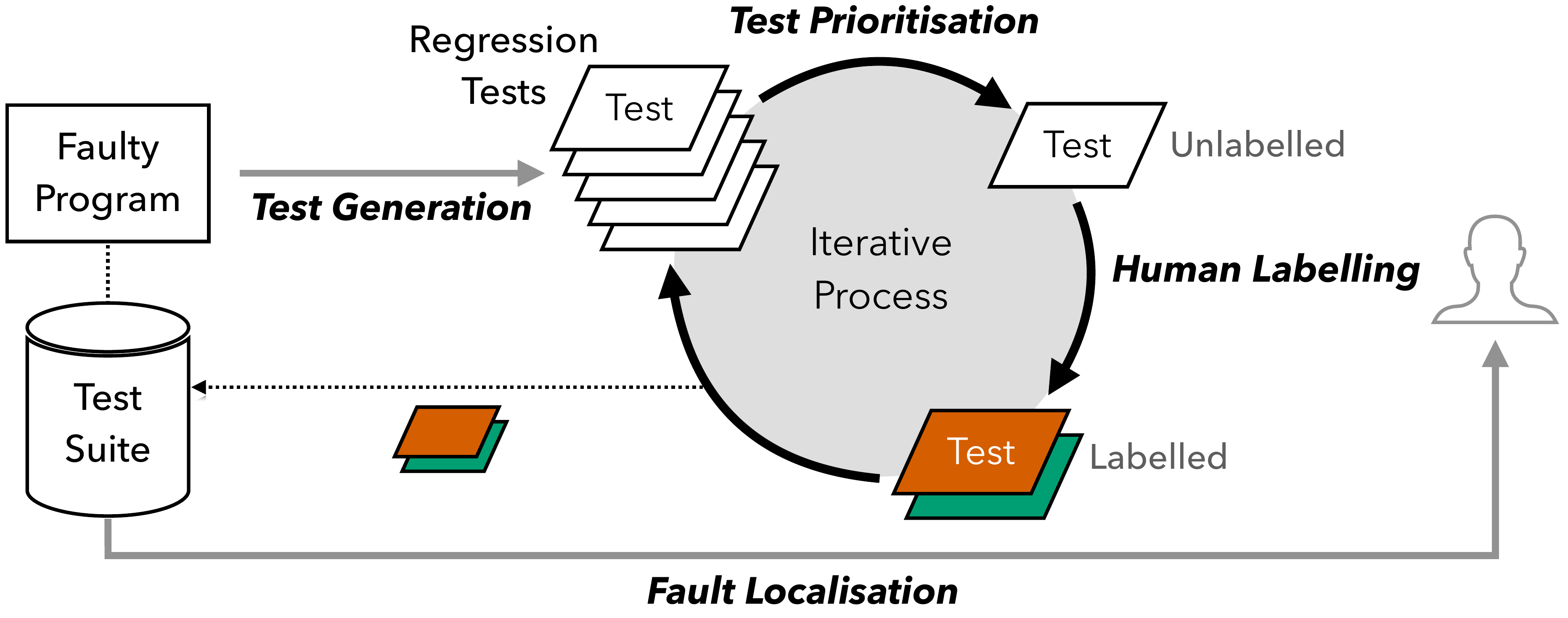}}
  \caption{Overview of Iterative Fault Localisation with Test Augmentation\label{fig:scenario}}
  \Description{Overview of Iterative Fault Localisation with Test Augmentation}
  \end{figure}

\subsection{Iterative Fault Localisation with Test Augmentation}
\label{sec:ifl_scenario}

Let us propose an Iterative Fault Localisation (IFL) scenario, in which \unifiedmetric is used to guide the augmentation of the test suite: an overview of the proposed scenario is presented in Fig.~\ref{fig:scenario}.

First, an automated test generation tool, such as \evo~\cite{fraser2011evosuite}
or Randoop~\cite{pacheco2007randoop}, produces regression tests for the
given faulty program (\textbf{Test Generation}): to reduce the number of 
generated test cases, we limit the scope of regression test generation only to 
the program elements covered by the initial failing test cases.
Since regression tests simply capture and record the current behaviour of the 
program, some assertions in the generated tests may capture the \textit{buggy}
behaviour of the program~\cite{pastore2013crowdoracles}. Note that 
the regression test cases generated by \evo always capture the behaviour 
of the current system as identity assertions~\cite{Fraser2012fr}.
Fig.~\ref{fig:evosuite_ex} shows an example of a test case automatically 
generated for the \dfj Math-59 buggy version by \evo: when calling 
\texttt{FastMath.max(0.0F, (-1653.0F))} (Line 3), the assertion generated from 
the buggy program expects the return value to be \texttt{-1653.0F} (Line 4), 
and not \texttt{0.0F}, which is the buggy behaviour of Math-59.

\begin{figure}[ht]
  \begin{lstlisting}
  @Test(timeout = 4000)
  public void test03()  throws Throwable  {
      float f0 = FastMath.max(0.0F, (-1653.0F));
      // wrong assertion! (f0 should be 0.0F)
      assertEquals((-1653.0F), f0, 0.01F);
  }
  \end{lstlisting}
  \caption{An \evo-generated test case for the class \texttt{FastMath} of Math-59 in Defects4J\label{fig:evosuite_ex}}
  \Description{An \evo-generated test case for the class \texttt{FastMath} of Math-59 in Defects4J}
\end{figure}

Consequently, unless the failure is detectable by implicit 
oracles~\cite{Barr:2015qd} such as program crashes or uncaught exceptions, 
a human engineer has to determine whether the program 
behaviour captured in the test case is \textit{correct} or \textit{incorrect}
(\textbf{Human Labelling}, or Oracle Elicitation). For example, an engineer 
will say that the test case in Fig.~\ref{fig:evosuite_ex} is capturing the 
\textit{incorrect} behaviour.
However, since manual oracle elicitation can be costly, tests should be 
presented to the engineer in the order of their relative diagnostic 
capabilities so that the more relevant tests to the fault localisation are 
labelled earlier. Therefore, at each iteration, all remaining test 
cases in the generated test suite are prioritised based on their diagnosability 
gain, and the best one is selected to be labelled by the engineer.
(\textbf{Test Prioritisation}). Note that the only cost not shown in
Fig.~\ref{fig:scenario} is that of measuring the coverage of generated tests,
which we expect to be automatable for \evo generated tests.

\begin{algorithm}[t]
\SetAlgoLined
\DontPrintSemicolon
\SetKwInput{kwPrecondition}{Precondition}
\SetCommentSty{mycommfont}
\KwIn{Faulty program $P$, Initial test suite $T$, Test results $R$,
Test generation tool \textsc{TestGenerator}, SBFL formula \textsc{FaultLocaliser}, Diagnostic capability metric $f$, Test generation time budget $b_t$, Querying budget $b_q$}
\kwPrecondition{$\exists t \in T. R(t) = 0$}
\KwOut{Fault localisation result}
 $P_{susp} \leftarrow \textsc{GetSuspiciousParts}(T, R)$ \\
 $T' \leftarrow \textsc{TestGenerator}(P_{susp}, b_t)$ \\
 $i \leftarrow 1$ \\
 \While{$i \leq  b_q \land T' \ne \emptyset$}{
  $t_s \leftarrow \argmax_{t \in T'} f(T, t)$\tcp*{select $t_s$ from $T'$}
  $R(t_s) \leftarrow \textsc{GetHumanLabel}(t_s)$\tcp*{label $t_s$}
  $T \leftarrow T \cup \{t_s\}$\tcp*{add $t_s$ to $T$}
  $T' \leftarrow T' \setminus \{t_s\}$\tcp*{remove $t_s$ from $T'$}
  $i \leftarrow i + 1$ \\
 }
 \Return{$\textsc{FaultLocaliser}(T,R)$}
 \caption{Iterative FL with Test Augmentation}
 \label{algo:ours}
\end{algorithm}

Algorithm~\ref{algo:ours} formally describes the workflow. The test generation 
tool, \textsc{TestGenerator}, generates tests $T'$ for the suspicious part $P_{susp}$ of a 
given faulty program $P$ within the time budget $b_t$ (Line 1-2). Then, 
until either the oracle querying budget, $b_q$, is exhausted or $T'$ becomes empty, a 
test case $t_s$ with the maximum diagnosability value is 
iteratively selected from from $T'$ (Line 4-5). If the test $t_s$ reveals buggy behaviour 
of a program, the user labels it as \textsc{Incorrect} (i.e., $R(t_s) = 0$), or 
otherwise as \textsc{Correct} (i.e., $R(t_s) = 1$) (Line 6). Once labelled, the test is 
moved from $T'$ to $T$ (Line 7-8). Finally, after the loop terminates, the 
final FL result is returned (Line 11).

\section{Experimental Setup}
\label{sec:exp_setup}

This section presents the research questions and describes the setup and the 
environment of our empirical evaluation.

\subsection{Research Questions}
\label{sec:rqs}

We ask the following research questions to evaluate both our newly proposed 
diagnosability gain metric, \unifiedmetric, and our IFL approach.

\subsubsection{RQ1. Effectiveness}
\textbf{How effective is \unifiedmetric 
when compared to the existing metrics?}
To answer RQ1, we use the same 
evaluation protocol used for existing metrics, and compare the results produced 
by \unifiedmetric to those produced by existing metrics. RQ1 is designed to 
investigate the performance of \unifiedmetric in an ideal situation with the 
perfect knowledge of coverage of all test cases as well as human-written test 
oracles. While some of the assumptions may be infeasible in practice, we aim to establish 
the upper bound of \unifiedmetric in an ideal setting, as a baseline to our 
subsequent evaluation of IFL scenario. We set $\alpha$ to 0.5 for \unifiedmetric.

In addition to mAP, we also report the \textbf{Top-n Accuracy} (acc@n), a 
widely-adopted measure of fault localisation performance. For all faulty 
subjects, acc@n counts the number of subjects where at least one of the faulty 
program elements are ranked within the top $n$ locations. As in mAP, we report 
acc@n after each iteration using the test cases chosen up to that point. When 
there are multiple test orderings for a single \dfj bug due to multiple failing 
test cases (see Section~\ref{sec:comparison_protocol}), we average all relevant 
acc@n values. 

\subsubsection{RQ2. IFL Performance}\textbf{How effectively does 
\unifiedmetric facilitate fault localisation by prioritising automatically 
generated test cases?}
RQ2 aims to investigate the performance of \unifiedmetric
when used in the IFL scenario described in 
Section~\ref{sec:ifl_scenario}. We conduct IFL as 
shown in Algorithm~\ref{algo:ours}, starting with an initial test suite, $T$, 
which contains only the \dfj-provided failing test cases. Our IFL scenario for 
RQ2 consists of the following elements:

\noindent\emph{\underline{Test Generation}}: We employ \evo~\cite{fraser2011evosuite} version
1.0.7\footnote{This is not an official release but is the most recent version 
on GitHub (commit \texttt{800e12}). For each fault, the maximum number of tests 
per class is set to 200, and the length of a test case is limited to $20$ to 
avoid too complex test cases being generated, which is difficult for engineers 
to investigate.} as a test data generation tool. The target of regression 
test generation is restricted to only the methods covered by initial failing 
test cases, as specified in the property \texttt{relevant.classes} in \dfj.
We use two time budgets ($b_t$), 3 and 10 minutes, and allocate the physical 
time budget to a class proportionally to the number of suspicious methods in 
the class. For example, consider suspicious classes A and B that contain five 
and ten suspicious methods, respectively. Under the 3-minute budget, we allocate
1 and 2 minutes to class A and B, respectively. We generate 10 test suites for 
each fault and time budget to cater for the randomness of \evo. 

\noindent\emph{\underline{Coverage \& FL}}: We measure the coverage achieved by generated 
test cases against all suspicious classes using Cobertura. We perform 
method level SBFL using Ochiai as described in 
Section~\ref{sec:comparison_protocol}.

\noindent\emph{\underline{Test Prioritisation}}: \unifiedmetric (with $\alpha = 0.5$) is 
used as the test prioritisation metric $f$ in Line 5 of 
Algorithm~\ref{algo:ours}.

\noindent\emph{\underline{Human Labelling}}: We \textit{simulate a perfect human oracle 
querying} by running the generated test cases on the fixed version. If a 
regression test case fails on the fixed version (due to oracle violation or 
compile error), we consider the test as a failing test case for the buggy 
version. We report results from $\{1, 3, 5, 10\}$ iterations by setting the 
oracle querying budget, $b_q$, accordingly. 

\subsubsection{RQ3. Robustness} \textbf{How robust is \unifiedmetric against 
human errors, when used to guide oracle querying?}
Our third research question 
concerns the assumption of the perfect human oracle because, in reality, the 
human engineer may make incorrect judgements about the generated oracles that 
capture the current behaviour of a buggy program. To study how robust our IFL 
scenario is against such mistakes, we simulate labelling errors by applying 
the probability $p \in \left\{0.1, 0.3, 0.5, 0.7, 0.9\right\}$ of flipping the 
perfect judgement, and observe how much the localisation performance 
deteriorates.

\subsubsection{RQ4. Parameter Tuning} \textbf{What is the ideal parameter $\alpha$ for \unifiedmetric?}
\unifiedmetric has a parameter $\alpha$ that adjusts the relative 
weights of \splitmetric and \covermetric. For previous research questions, 
$\alpha$ has been set to 0.5 to assume equal contribution from \splitmetric and 
\covermetric to the measure of diagnosability gain and, consequently, to the 
prioritisation. Our final research question studies the impact of the parameter 
$\alpha$ on the performance of \unifiedmetric. To answer RQ4, we perform a grid 
search for $\alpha \in \{0.1, 0.3, 0.5, 0.7, 0.9\}$ to find its value that 
leads to the best prioritisation performance for \evo-generated tests.

\begin{table}[ht]
\caption{Experimental Subject - \dfj}
\label{tab:subject}
\begin{center}
\scalebox{0.78}{
\begin{tabular}{c|r|r|r|r|r|r|r}
\toprule
\multirow{3}{*}{Subject} & \multirow{2}{*}{\# Faults} & \multirow{2}{*}{kLoC} & \multicolumn{2}{c|}{Avg. \# Tests} & \multicolumn{3}{c}{Avg. \# Methods} \\ \cmidrule{4-8}
& & &Total& Failing& Total &\makecell{Suspi\\cious}& Faulty \\
\midrule
Commons-\textbf{lang} & 65 & 22 & 1,527 & 1.9 & 2,245 & 14.7 & 1.5\\
Commons-\textbf{math} & 106 & 85 & 2,713 & 1.7 & 3,602 & 53.2 & 1.7\\
JFree\textbf{Chart} & 26 & 96 & 4,903 & 3.5 & 2,205 & 127.5 & 4.5\\
Joda-\textbf{Time} & 27 & 28 & 1,946 & 2.8 & 4,130 & 375.4 & 2.0\\
\textbf{Closure} compiler & 133 & 90 & 5,038 & 2.6 & 7,927 & 855.4 & 1.8\\
\midrule
Total & 357 & \multicolumn{6}{c}{}\\
\bottomrule
\end{tabular}}
\end{center}
\end{table}

\subsection{Subject}
\label{sec:subjects}

Our empirical evaluation considers 357 faults in five different 
projects provided by \dfj version 2.0.0~\cite{just2014defects4j}, a real-world 
fault benchmark of Java programs. Each fault in \dfj is in the program source 
code, not in the configuration nor test files, and the corresponding patch is 
provided as a \textit{fixing commit}. For each faulty program, human-written 
test cases, at least one of which is bound to fail due to the fault, are 
provided. The failing tests all pass once the fixing commit is applied to the 
faulty version. Table~\ref{tab:subject} shows the statistics of our subjects. 
The \textit{Suspicious} column contains the average number of methods covered 
by at least one failing test.

For RQ1, we exclude a total of six faults from the 357 in Table~\ref{tab:subject}: two omission faults, Lang-23 and Lang-56, as they cannot be localised within the faulty version by SBFL techniques, as well as four deprecated faults (Lang-2, Time-21, Closure-63, and Closure-93).

For RQ2 to RQ4, we exclude two additional faults, Time-5 and Closure-105, 
because the \dfj-provided lists of relevant (suspicious) classes of those 
faults do not contain the actual faulty class due to an unknown reason. 
Overall, 349 faults are used, with 2.3 failing test cases on average per fault 
in the initial test suites.

\subsection{Implementation \& Environment}

All our experiment have been performed on machines equipped with
Intel Core i7-7700 CPU and 32GB memory, running Ubuntu 16.04.
We use Java 8 for \evo and Cobertura.
Our replication package, available online at
\url{https://github.com/agb94/FDG-artifact}~\cite{fdg-artifact}, includes all implementation
as well as the data required to reproduce our work.

\section{Results}
\label{sec:results}

This section presents the answers to our research questions based on the results of our empirical evaluations.

\begin{table*}[t]
  \caption{The acc@n values with the selected human-written test cases of \dfj at each iteration
  (\# total subjects = 351)}
  \begin{center}
  \scalebox{0.81}{
    \renewcommand{\arraystretch}{1.3}
    \begin{tabular}{c|c|rrrr|rrrr|rrrr|rrrr|rrrr|rrrr}
\toprule
\multicolumn{2}{c|}{} & \multicolumn{24}{c}{acc}\\\cmidrule{3-26}
\multicolumn{2}{c|}{} &   @1 &   @3 &   @5 &   @10 &   @1 &   @3 &   @5 &   @10 &   @1 &   @3 &   @5 &   @10 &   @1 &   @3 &   @5 &   @10 &   @1 &   @3 &   @5 &   @10 &   @1 &   @3 &   @5 &   @10\\
   \midrule\midrule
   \multicolumn{2}{c|}{Metric} & \multicolumn{4}{c|}{EntBug (mAP=0.119)} & \multicolumn{4}{c|}{Total (mAP=0.119)} & \multicolumn{4}{c|}{DDU (mAP=0.149)} & \multicolumn{4}{c|}{Add. (mAP=0.158)} & \multicolumn{4}{c|}{RAPTER (mAP=0.158)} & \multicolumn{4}{c}{TfD (mAP=0.208)}\\   \midrule
\multirow{12}{*}{Iter.} & Init. & 4        & 30        & 45        &   65 & 4        & 30        & 45        &   65 & 4        & 30        & 45        &   65 & 4        & 30        & 45        &   65 & 4        & 30        & 45        &   65 & 4        & 30        & 45        &   65 \\\cmidrule{2-26}
& 1           &   9 &  40 &  61 &   80 &   9 &  40 &  61 &   80 &  10 &  43 &  64 &   85 &   9 &  39 &  60 &   80 &   9 &  39 &  60 &   80 &  21 &  51 &  67 &   84 \\
& 2           &   9 &  41 &  63 &   85 &   9 &  41 &  63 &   85 &  12 &  49 &  74 &   99 &  10 &  44 &  67 &   94 &  10 &  44 &  67 &   94 &  28 &  67 &  81 &  100 \\
& 3           &   9 &  40 &  64 &   90 &   9 &  40 &  64 &   90 &  14 &  53 &  76 &  107 &  11 &  47 &  71 &   99 &  12 &  48 &  72 &  100 &  32 &  71 &  88 &  112 \\
& 4           &   9 &  40 &  64 &   94 &   9 &  40 &  64 &   94 &  15 &  54 &  77 &  112 &  12 &  52 &  76 &  104 &  13 &  53 &  76 &  104 &  33 &  81 &  95 &  127 \\
& 5           &   9 &  41 &  67 &   95 &   9 &  41 &  67 &   95 &  15 &  54 &  79 &  115 &  12 &  54 &  80 &  112 &  13 &  53 &  81 &  112 &  35 &  82 &  98 &  130 \\
& 6           &   9 &  41 &  67 &   96 &   9 &  41 &  67 &   96 &  15 &  55 &  79 &  118 &  14 &  56 &  82 &  115 &  16 &  56 &  83 &  117 &  35 &  87 & 100 &  134 \\
& 7           &   9 &  41 &  68 &   96 &   9 &  41 &  68 &   96 &  15 &  56 &  81 &  123 &  15 &  57 &  85 &  119 &  16 &  57 &  84 &  117 &  36 &  89 & 105 &  142 \\
& 8           &   9 &  41 &  68 &   99 &   9 &  41 &  68 &   99 &  15 &  56 &  81 &  125 &  17 &  60 &  86 &  121 &  18 &  59 &  85 &  123 &  37 &  89 & 109 &  145 \\
& 9           &  11 &  43 &  68 &   99 &  11 &  43 &  68 &   99 &  16 &  58 &  82 &  127 &  19 &  62 &  87 &  125 &  20 &  61 &  87 &  126 &  37 &  91 & 111 &  149 \\
& 10          &  12 &  44 &  68 &  101 &  12 &  44 &  68 &  101 &  17 &  59 &  84 &  127 &  19 &  65 &  88 &  127 &  20 &  63 &  87 &  127 &  39 &  93 & 111 &  149 \\\cmidrule{2-26}
& Full & 110 & 208 & 250 &  277 & 110 & 208 & 250 &  277 & 110 & 208 & 250 &  277 & 110 & 208 & 250 &  277 & 110 & 208 & 250 &  277 & 110 & 208 & 250 &  277\\
\midrule \midrule
   \multicolumn{2}{c|}{Metric} & \multicolumn{4}{c|}{FLINT (mAP=0.113)} & \multicolumn{4}{c|}{Prox (mAP=0.250)} & \multicolumn{4}{c|}{S3 (mAP=0.271)} & \multicolumn{4}{c|}{\textbf{\splitmetric} (mAP=0.277)} & \multicolumn{4}{c|}{\textbf{\covermetric} (mAP=0.278)} & \multicolumn{4}{c}{\textbf{\unifiedmetric} (mAP=\textbf{0.298})} \\   \midrule
    \multirow{12}{*}{Iter.}   & Init. & 4        & 30        & 45        &   65 & 4        & 30        & 45        &   65 & 4        & 30        & 45        &   65 & 4        & 30        & 45        &   65 & 4        & 30        & 45        &   65 & 4        & 30        & 45        &   65 \\\cmidrule{2-26}
    & 1           & 4       & 31       & 47       & 69       & \textbf{27} & 71       & \textbf{97} & \textbf{130} & 22       & 51       & 69       & 92       & 22      & 51       & 69       & 92       & 25       & \textbf{75}  & 94        & 122      & 24       & 64        & 87        & 109       \\
    & 2           & 8       & 45       & 66       & 79       & 34       & 87       & 118      & 145       & 32       & 70       & 93       & 124      & 31      & 70       & 92       & 121      & 35       & 93        & 113       & 145      & \textbf{41} & \textbf{100} & \textbf{124} & \textbf{155} \\
    & 3           & 8       & 47       & 68       & 79       & 40       & 97       & 122      & 158       & 40       & 88       & 108      & 137      & 40      & 91       & 112      & 146      & 46       & 105       & 130       & 164      & \textbf{48} & \textbf{107} & \textbf{135} & \textbf{175} \\
    & 4           & 8       & 48       & 69       & 81       & 42       & 107      & 130      & 168       & 47       & 97       & 124      & 151      & 46      & 106      & 126      & 153      & \textbf{53} & 114       & \textbf{145} & 181      & 52       & \textbf{115} & \textbf{145} & \textbf{190} \\
    & 5           & 8       & 48       & 70       & 82       & 45       & 109      & 137      & 174       & 54       & 105      & 128      & 158      & 52      & 114      & 135      & 160      & 54       & \textbf{121} & 154       & 188      & \textbf{57} & \textbf{121} & \textbf{158} & \textbf{204} \\
    & 6           & 9       & 49       & 72       & 83       & 47       & 112      & 140      & 182       & 56       & 113      & 135      & 160      & 59      & 122      & 141      & 164      & 55       & 128       & 155       & 192      & \textbf{60} & \textbf{130} & \textbf{164} & \textbf{212} \\
    & 7           & 9       & 49       & 72       & 86       & 47       & 115      & 144      & 185       & 63       & 118      & 141      & 164      & 60      & 125      & 144      & 171      & 57       & 131       & 162       & 198      & \textbf{64} & \textbf{135} & \textbf{170} & \textbf{221} \\
    & 8           & 9       & 50       & 73       & 86       & 48       & 120      & 153      & 188       & \textbf{66} & 120      & 146      & 171      & 62      & 129      & 151      & 176      & 61       & 133       & 167       & 204      & 65       & \textbf{139} & \textbf{174} & \textbf{219} \\
    & 9           & 11      & 52       & 75       & 86       & 49       & 122      & 155      & 190       & \textbf{66} & 124      & 149      & 179      & 64      & 130      & 153      & 178      & 62       & 133       & 175       & 209      & \textbf{66} & \textbf{145} & \textbf{178} & \textbf{220} \\
    & 10          & 11      & 52       & 75       & 86       & 50       & 125      & 160      & 195       & \textbf{70} & 127      & 153      & 180      & 68      & 134      & 157      & 189      & 63       & 139       & 170       & 212      & 68       & \textbf{148} & \textbf{179} & \textbf{222} \\\cmidrule{2-26}
    & Full & 110 & 208 & 250 &  277 & 110 & 208 & 250 &  277 & 110 & 208 & 250 &  277 & 110 & 208 & 250 &  277 & 110 & 208 & 250 &  277 & 110 & 208 & 250 &  277\\
\bottomrule 
\end{tabular}
  }
  \label{tab:RQ1}
  \end{center}
\end{table*}

\subsection{RQ1: Effectiveness}
\label{sec:rq1:results}

Table~\ref{tab:RQ1} shows the localisation performance of each metric on
\dfj human-written tests. Each row presents acc@n values ($n \in [1,3,5,10]$)
of SBFL results at each iteration; rows \textit{Init.} and \textit{Full}
represent the results when using a single initial failing test case and full 
test suite (about 4.2K tests on average), respectively. The numbers in bold 
represent the highest values in the iteration for the corresponding $n$ value. 
Additionally, the mAP values after ten iterations are shown next to the name of each metric.

The results show that \unifiedmetric outperforms all nine studied metrics by 
achieving the highest acc@3, 5, 10 values in all iterations except for the 
first. Especially at the tenth iteration, acc@10 is 23\% and 14\% higher than 
those obtained by the state-of-the-art metrics, S3 and Prox, respectively. 
Furthermore, we report the performance of \splitmetric and \covermetric as 
stand-alone metrics. Although they do not perform as well as \unifiedmetric, the two 
subcomponents still achieve higher mAP values than all existing metrics after 
ten iterations, showing each of them can outperform the existing metrics as a 
stand-alone diagnosability metric.

Additionally, we also compare the FL accuracy at the tenth iteration to the 
accuracy obtained using the initial test set only, as well as the full test 
suites. After adding ten test cases based on \unifiedmetric, acc@1 and acc@10 
values are increased by 17.0 and 3.4 times, respectively, compared to the 
initial test set. Notably, \unifiedmetric achieves $62\%$ of acc@1 and $80\%$ 
of acc@10 compared to the full test suite, after ten iterations, despite the 
fact that the full test suites have approximately 380 times (=4.2K/11) more 
tests than the 11 test cases obtained using \unifiedmetric (one initial + ten 
additional).

\begin{tcolorbox}[boxsep=2pt,left=2pt,right=2pt,top=1pt,bottom=1pt]
\textbf{Answer to RQ1:} \unifiedmetric significantly outperforms all nine 
studied metrics achieving the highest acc@n values at almost every iteration. 
When ten additional test cases are selected, \unifiedmetric shows at least 14\% 
higher acc@10 compared to other metrics. 
\end{tcolorbox}

\subsection{RQ2: IFL Performance}
\label{sec:rq2:results}
\begin{table}[ht]
\caption{Average number of tests generated by \evo
(number of failing tests in parenthesis)}
\label{tab:RQ2_stat}
\begin{center}
\scalebox{0.88}{
  \begin{tabular}{c|r|r|r|r|r}
\toprule
Time &  \multicolumn{5}{c}{Project}\\ \cmidrule{2-6}
Budget & \multicolumn{1}{c|}{Lang} & \multicolumn{1}{c|}{Math} & \multicolumn{1}{c|}{Chart} & \multicolumn{1}{c}{Time} & \multicolumn{1}{c}{Closure} \\ \midrule
3 mins & 32 (1.0) & 97 (0.8) & 225 (2.4) & 442 (0.7) & 832 (0.2)\\
10 mins & 34 (1.1) & 102 (0.9) & 233 (2.5) & 475 (0.8) & 964 (0.4)\\
\bottomrule
\end{tabular}}
\end{center}
\end{table}

Table~\ref{tab:RQ2_stat} shows the average number of total and failing test 
cases generated by \evo for the studied faults, using 3 and 10 minutes 
as time budgets, $b_t$, respectively (hereafter denoted by T3 and T10). These tests 
correspond to $T'$ in Algorithm~\ref{algo:ours}. Using these 
generated test cases, we evaluate fault localisation performance after a 
different number of oracle queries. 

\begin{figure}[t]
  \centerline{\includegraphics[width=\linewidth]{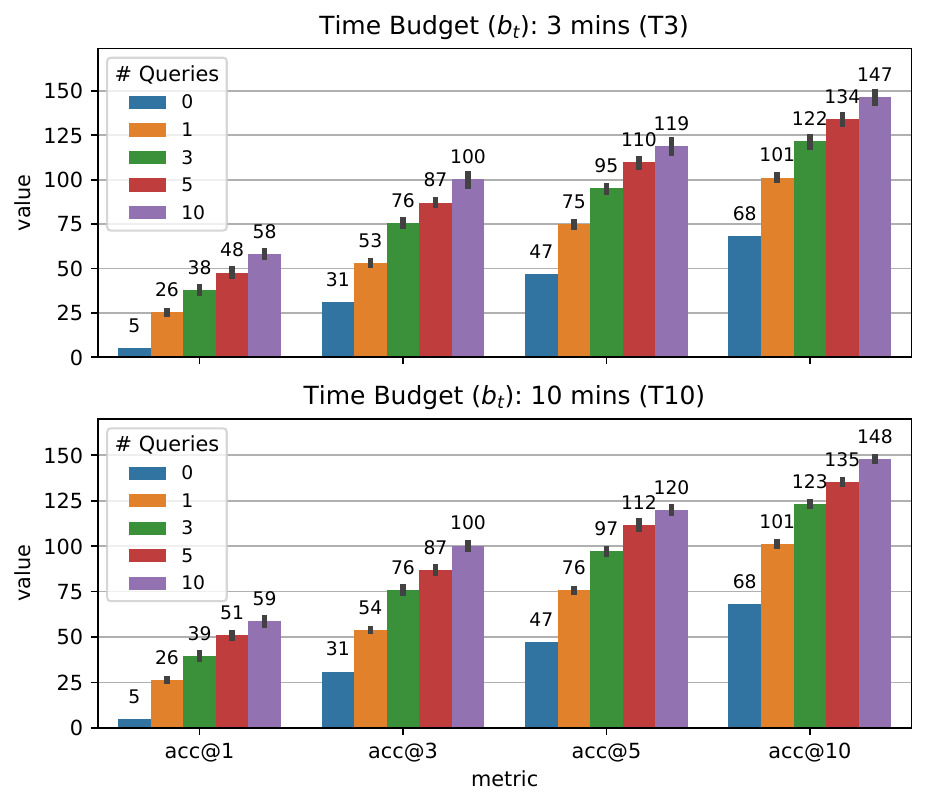}}
  \caption{The acc@n values for each time and query budgets. The error bars show
  the standard error from different seeds. (\# total subjects = 349)}
  \label{fig:RQ2}
\end{figure}

Fig.~\ref{fig:RQ2} shows how the acc@n values change as the query budget 
increases for T3 (top) and T10 (bottom) scenarios. Overall, T10 has slightly 
better localisation results and less standard deviation in performance than T3, 
although the difference is not significant. This is as expected, for 
T10 is larger, thus more likely to contain diverse test cases.
We note that the effectiveness of test suite augmentation 
with automatically generated test cases is less than that of augmentation with 
human-written test cases. However, augmenting the test suite with generated 
test cases also increases the SBFL accuracy significantly. The test suite with
ten newly labelled test cases achieves 11.6x acc@1 ($=58/5$) and 2.2x acc@10 
($=147/68$) compared to those of the initial test suite (\# queries = 0).

Interestingly, we note that accurate fault localisation does not always require 
many additional \textit{failing} test cases to be generated. The localisation 
results tend to be positive despite the small number of failing test cases that 
have been generated (see Table~\ref{tab:RQ2_stat}). This is because generated 
passing test cases can still effectively contribute to decreasing the 
suspiciousness of non-faulty program elements.

\begin{tcolorbox}[boxsep=2pt,left=2pt,right=2pt,top=1pt,bottom=1pt]
\textbf{Answer to RQ2:} Given a small set of failing test cases only, 
the accuracy of SBFL can be greatly improved by querying human engineers about 
the oracles for a small number of automatically generated tests prioritised by 
\unifiedmetric.
\end{tcolorbox}

\begin{figure}[t]
  \centerline{\includegraphics[width=\linewidth]{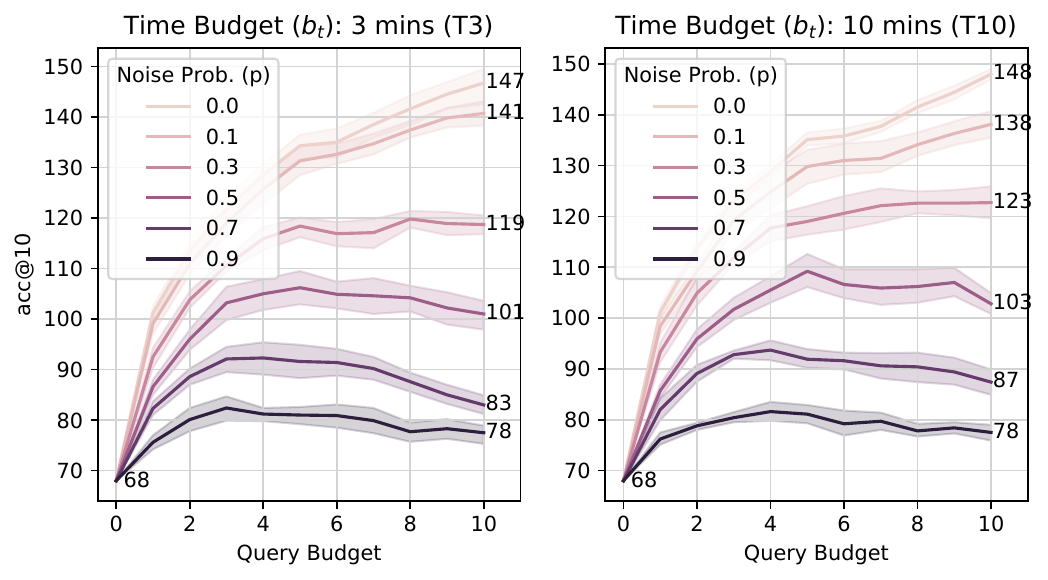}}
  \Description{The acc@10 values for each labelling error rate and query budget.
  Each band shows the standard error of the observations from different seeds.}
  \caption{The acc@10 values for each labelling error rate and query budget.
  Each band shows the standard error of the observations from different seeds.}
  \label{fig:RQ3}
\end{figure}

\subsection{RQ3: Robustness}
\label{sec:rq3:results}

Fig.~\ref{fig:RQ3} shows how localisation performance varies against 
different labelling error rates. The results show that localisation 
performance, acc@10, gradually deteriorates as the error rate ($p$) 
increases. However, we observe that the localisation results are fairly 
robust up to the error rate of $p = 0.3$: the acc@10 values still increase as 
the query budget increases. Based on the report from Pastore et 
al.~\cite{pastore2013crowdoracles} that shows a qualified crowd can achieve 
accuracy above 0.69 when classifying incorrect assertions for \evo-generated 
test cases, the error rate tolerance of up to 0.3 shows the robustness of 
\unifiedmetric.

Overall, the results show that, despite some errors in labelling, adding new 
test cases significantly increases the SBFL accuracy compared to the 
initial test suite. This is because the correctly labelled subsequent test  
cases can mitigate the impact of earlier labelling errors.
Further, we observe that T10 is slightly more robust than T3 against the same 
error rate. This suggests that the richer the generated test suite is,
the more likely it contains tests that can adjust the errors caused by the 
incorrect labels.

\begin{tcolorbox}[boxsep=2pt,left=2pt,right=2pt,top=1pt,bottom=1pt]
\textbf{Answer to RQ3:} Although SBFL performance deteriorates as the 
labelling error rate increases, the results show that \unifiedmetric can be 
resilient to labelling errors, especially up to the error rate of 30\%. A 
richer and larger test suites are also more resilient to labelling errors than 
smaller test suites.
\end{tcolorbox}

\subsection{RQ4: Parameter Tuning}
\label{sec:rq4:results}

Table~\ref{tab:RQ4} shows the performance of \unifiedmetric obtained while 
varying $\alpha$ from 0.1 to 0.9 with the interval of 0.2: a value higher than 
0.5 means that \unifiedmetric puts more emphasis on \splitmetric, while lower 
than 0.5 puts more emphasis on \covermetric. When only one query is made, FL 
performance measured using acc@10 is best at $\alpha = 0.5$. However, as more
queries are made, the higher $\alpha$ values, e.g. 0.7, 0.9, or 1.0, tend to
produce better performance. 
These results show that it is better to balance the weights of two metrics ($\alpha$ = 0.5)
when the query budget is low, whereas it is better to focus on 
splitting the ambiguity groups ($\alpha > 0.5$) when there are multiple chances
to query test oracles from developers.
Since both program structure and test suite composition can significantly 
affect the ambiguity groups, we suggest that $\alpha$ needs to be tuned based 
on actual fault data when used in practice. However, the results also show that 
\unifiedmetric can outperform state-of-the-art diagnosability metrics even with
the default value of $\alpha=0.5$.

\begin{tcolorbox}[boxsep=2pt,left=2pt,right=2pt,top=1pt,bottom=1pt]
\textbf{Answer to RQ4:} When oracle querying is performed only once, FL
performance increases the most when using \unifiedmetric with $\alpha=0.5$.
However, as more queries
are made, the higher $\alpha$ values lead to higher FL
performance improvement.
\end{tcolorbox}

\begin{table}[t]
  \caption{The averaged acc@10 values for T10 test suites}
  \label{tab:RQ4}
  \centering
    \scalebox{0.88}{ 
    \renewcommand{\arraystretch}{1.3}
    \begin{tabular}{l|r|rrrr}
    \toprule
       \multirow{2}{*}{Diagnosability Metric} & \multicolumn{5}{c}{\# Queries}\\\cmidrule{2-6}
               &  0 &    1  &     3  &     5  &     10 \\
    \midrule
   FDG ($\alpha=0.0$) = \covermetric & 68 &          98.4 &  121.2 &  131.4 &  144.0 \\
   FDG ($\alpha=0.1$) & 68 &           98.7 &           122.1 &         131.8 &  144.9 \\
   FDG ($\alpha=0.3$) & 68 &           98.7 &           125.3 &         135.9 &  146.6 \\
   FDG ($\alpha=0.5$) & 68 & \textbf{101.1} &           123.1 &         135.1 &  148.0 \\
   FDG ($\alpha=0.7$) & 68 &           98.1 &  \textbf{126.1} &         135.9 &  148.7 \\
   FDG ($\alpha=0.9$) & 68 &           93.3 &           125.7 &         135.3 &  \textbf{151.7} \\
   FDG ($\alpha=1.0$) = \splitmetric & 68 &          91.8 &  120.7 &  \textbf{139.8} &  151.0 \\\midrule\midrule
                Prox  & 68 &           99.5 &           119.7 &         126.6 &  137.8 \\
                S3    & 68 &           91.8 &           117.3 &         133.9 &  146.3 \\
    \bottomrule
  \end{tabular}}
\end{table}

\section{Discussion}
\label{sec:discussion}

We discuss some assumptions and implications related to \unifiedmetric.

\subsection{Rankings and FL}
\label{sec:rankings}

Whether reporting in the form of linear rankings is the best vessel for fault 
localisation remains under debate: Parnin and Orso state that human developers do 
not really follow a given ranking linearly while 
debugging~\cite{parnin2011automated}, while Xia et al. state that they do help, 
especially if they are of high quality~\cite{Xia2016fi}. In this work, we 
simply adopt the ranking based evaluation as one possible way of quantitatively 
measuring the diagnosability gain. 

Note that higher SBFL rankings mean more diverse patterns and fewer ambiguity 
groups in the achieved coverage, both of which contribute to the general 
diagnosability of a test suite. In fact, it is entirely possible that the human 
labelling phase (see Section~\ref{sec:ifl_scenario}) can be the debugging 
activity itself, as the labeller will have to consider program behaviour and 
test results for the given task. However, even in such a scenario, we expect 
that \unifiedmetric will still prioritise test results efficiently for the 
labeller. Such a human study would be a very interesting future work.

\subsection{Retaining Generated Test Cases}
\label{sec:retention}

The test suite augmentation scenario proposed in Section~\ref{sec:ifl_scenario} 
assumes that 1) there is no other test case apart from the failing test cases, 
and 2) the new test cases are generated for the purpose of FL only. What if 
there already exists a test suite, but the newly generated test cases can 
further improve its diagnosability? The generated test cases do not have to be 
single-use only: once labelled, selected test cases can be retained to be part 
of the existing test suites, as long as the testing budget can afford them. 

\subsection{\unifiedmetric as Test Generation Objective}
\label{sec:objective}

Instead of prioritising already generated test cases using \unifiedmetric, it 
may also be possible to use to directly adopt \unifiedmetric as a fitness (or 
an objective) function that guides test case generation. Note that 
\unifiedmetric itself can be considered as a weighted-sum of \covermetric 
and \splitmetric, which will try to achieve coverage and break ambiguity groups, 
respectively. As such, we expect that search-based test data generation guided 
by \unifiedmetric may yield test suites with high coverage and high 
diagnosability.

\section{Related Work}
\label{sec:relatedwork}

We have discussed existing work on diagnosability metrics for test cases 
in Section~\ref{sec:survey}. This section further discusses fault 
localisation techniques that involve human-in-the-loop, as well as a test suite 
augmentation for fault localisation and algorithmic debugging, which are 
related to our IFL scenario in 
Section~\ref{sec:ifl_scenario}.

\subsection{Human-in-the-Loop Debugging}
Many debugging techniques depend on user feedback. Types of required feedback 
include the correctness of a specific program 
element~\cite{bandyopadhyay2012tester,gong2012interactive}, the correctness of 
a specific variable value~\cite{lin2017feedback}, the correctness of a specific 
execution~\cite{li2016iterative}, and the relative location of the fault with 
respect to breakpoints~\cite{hao2009interactive}; some techniques provide 
visual feedback~\cite{hao2009vida,ko2004designing, ko2008debugging}. 
\unifiedmetric requires the user to check the automatically generated test 
assertions, which is similar to checking the correctness of 
executions~\cite{li2016iterative}. However, the main contribution of 
\unifiedmetric lies in the way it chooses which test to get the feedback about. 
ENLIGHTEN~\cite{li2018enlightened} asks the developer to investigate 
input-output pairs of the most suspicious method invocations: the feedback is 
then encoded as a virtual test and used to update the localisation results. 
Recently, B{\"o}hme et al. proposed Learn2Fix~\cite{bohme2019human}, a 
human-in-the-loop program repair technique. Learn2Fix generates test cases by 
mutating the failing test and allowing the user to label them in the order of 
their likelihood of failing. Learn2Fix trains an automatic bug oracle using 
user feedback, which is in turn used to amplify the test suite for better patch 
generation. While our iterative localisation scenario also requires human 
interaction, the scenario differs from existing techniques in that it only 
requires judgements on the correctness of automatically generated assertions. 

\subsection{Test Augmentation for Fault Localisation}

Automated test generation has been widely used to support fault localisation by 
augmenting insufficient test suites. Artzi et al.~\cite{artzi2010directed} used 
dynamic symbolic execution to generate test cases that are similar to failing 
executions~\cite{renieres2003fault}. This type of methodology also can be 
plugged into the part of our debugging scenario as a test generation tool. 
BUGEX~\cite{robetaler2012isolating} generates additional test cases similar to 
a given failing one using \evo and uses an automated oracle to differentiate 
between passing and failing executions. Finally, BUGEX identifies the runtime 
properties that are relevant to the failure by comparing passing and failing 
executions. Compared to BUGEX, our debugging scenario does not assume the 
existence of an automated bug oracle and instead uses test prioritisation to 
reduce the cost of querying oracles from a developer.
$F^3$~\cite{jin2013f3} is a fault localisation technique for field failures, 
which extends a bug reproduction technique, BugRedux~\cite{jin2012bugredux}: 
it synthesises failing and passing executions similar to the field failure and 
uses them for its customised fault localisation technique. While $F^3$
can only debug program crashes that can be detected implicitly,
our scenario aims to localise faults where no such automatic oracle is 
available.

Xuan et al.~\cite{xuan2014test} split test cases into smaller test cases to 
increase the fault diagnosability of the given test suite. In comparison, we 
consider cases in which only a few failing test cases are available and use 
an automated test data generation technique to support fault localisation. 
Recently, Kuma et al.~\cite{kuma2020improving} presented several strategies for 
selecting the minimum number of test cases out of many automatically generated 
test cases. However, they only focus on differences in a program spectrum and 
do not quantify the diagnosability of test cases, which is why we 
exclude this work from our analysis. Moreover, Kuma et al. use the behaviour of 
the past version as the test oracle, which may limit its applicability due to 
limited access to the past versions or version compatibility issues.

\subsection{Algorithmic Debugging}

Algorithmic debugging~\cite{Caballero2017rz} was first proposed by 
Shapiro~\cite{Shapiro1982jk} for logic programming languages such as PROLOG. 
Shapiro introduces an approach that can systemically narrow down the cause of 
incorrect computation using what is called a debugging tree, whose root node 
corresponds to the final outcome of the computation, while each subtree 
represents an intermediate computation whose result is used to compute its 
parent node. The debugger can systematically guide the user through the 
debugging tree, narrowing down the possible root cause of the incorrect 
computation. \unifiedmetric shares a few similarities with algorithmic 
debugging: the user is expected to judge the correctness of the final 
computation (i.e., the test oracle), and the technique tries to systematically 
narrow down the root cause by breaking ambiguity groups. However, 
\unifiedmetric does not require the user to make any judgement about any target 
program elements, allowing anyone who understands the input-output 
specification to use it.

\section{Threats to Validity}
\label{sec:threats}

Threats to internal validity regard factors that may influence the observed 
effects, such as the integrity of the coverage and test results data, as well 
as the test data generation and fault localisation. To mitigate such threats, 
we have used the widely studied Cobertura and \evo, as well as the publicly 
available scripts in \dfj, to collect or generate data.

Threats to external validity concern any factors that may limit the 
generalisation of our results. Our results are based on \dfj, a 
benchmark against which many fault localisation techniques are evaluated. 
However, only further experimentations using more diverse subject programs and 
faults can strengthen the generalisability of our claims. We also consider an iterative scenario, in which tests are chosen and added one by one. 
It is possible that non-constructive heuristics such as Genetic Algorithm
may produce different and potentially better orderings. Our approach 
is based on the assumption that an iterative process will make it easier for the human engineer to make the oracle judgement.

Finally, threats to construct validity concern situations where used metrics 
may not reflect the actual properties they claim to measure. All evaluation 
metrics used in our study are widely used in fault localisation literature, 
leaving little room for misunderstanding. It is possible that Coincidental Correctness (CC)~\cite{masri2009empirical} has interfered with our measurements, as it is known to exist in \dfj~\cite{abou2019coincidental}. However, it is theoretically impossible to entirely filter out CC from test results. We also note that all coverage-based fault localisation techniques are equally affected by CC.

\section{Conclusion}
\label{sec:conclusion}

We propose \unifiedmetric, a novel measure of the diagnosability of test 
cases for SBFL. When evaluated with the human-written test cases in \dfj, 
\unifiedmetric can successfully augment single failing test cases, so that 
adding only ten more test cases improves acc@10 by 3.4 times. The augmentation 
based on \unifiedmetric also achieves 23\% and 14\% higher acc@10 when compared 
to those based on state-of-the-art metrics, S3 and Prox, respectively. 
We also introduce an iterative fault localisation scenario, in which the 
localisation starts with insufficient test suites that contain only failing 
test cases. By automatically generating test cases using \evo, and prioritising 
them for human oracle judgements using \unifiedmetric, we show that acc@1 can 
be improved by 11.6 times after only ten human interactions. Future work will 
consider closer integration of \unifiedmetric and test data generation to 
improve the overall efficiency of our approach. In particular, we expect that 
\unifiedmetric can provide guidance for search based test data generation to 
improve the the diagnosability of the generated test cases.

\section*{Acknowledgment}
Gabin An and Shin Yoo are supported by National Research Foundation of Korea 
(NRF) Grant (NRF-2020R1A2C1013629), Institute for Information \& 
communications Technology Promotion grant funded by the Korean government 
(MSIT) (No.2021-0-01001), and Samsung Electronics (Grant No. IO201210-07969-01).

\bibliographystyle{ACM-Reference-Format}
\bibliography{references}


\begin{thebibliography}{59}

%%% the Web address.

\ifx \showCODEN    \undefined \def \showCODEN     #1{\unskip}     \fi
\ifx \showDOI      \undefined \def \showDOI       #1{#1}\fi
\ifx \showISBNx    \undefined \def \showISBNx     #1{\unskip}     \fi
\ifx \showISBNxiii \undefined \def \showISBNxiii  #1{\unskip}     \fi
\ifx \showISSN     \undefined \def \showISSN      #1{\unskip}     \fi
\ifx \showLCCN     \undefined \def \showLCCN      #1{\unskip}     \fi
\ifx \shownote     \undefined \def \shownote      #1{#1}          \fi
\ifx \showarticletitle \undefined \def \showarticletitle #1{#1}   \fi
\ifx \showURL      \undefined \def \showURL       {\relax}        \fi
% invisible to TeX
\providecommand\bibfield[2]{#2}
\providecommand\bibinfo[2]{#2}
\providecommand\natexlab[1]{#1}
\providecommand\showeprint[2][]{arXiv:#2}

\bibitem[\protect\citeauthoryear{Abou~Assi, Trad, Maalouf, and Masri}{Abou~Assi
  et~al\mbox{.}}{2019}]%
        {abou2019coincidental}
\bibfield{author}{\bibinfo{person}{Rawad Abou~Assi}, \bibinfo{person}{Chadi
  Trad}, \bibinfo{person}{Marwan Maalouf}, {and} \bibinfo{person}{Wes Masri}.}
  \bibinfo{year}{2019}\natexlab{}.
\newblock \showarticletitle{Coincidental correctness in the Defects4J
  benchmark}.
\newblock \bibinfo{journal}{\emph{Software Testing, Verification and
  Reliability}} \bibinfo{volume}{29}, \bibinfo{number}{3}
  (\bibinfo{year}{2019}), \bibinfo{pages}{e1696}.
\newblock


\bibitem[\protect\citeauthoryear{Abreu, Zoeteweij, and Van~Gemund}{Abreu
  et~al\mbox{.}}{2006}]%
        {abreu2006evaluation}
\bibfield{author}{\bibinfo{person}{Rui Abreu}, \bibinfo{person}{Peter
  Zoeteweij}, {and} \bibinfo{person}{Arjan~JC Van~Gemund}.}
  \bibinfo{year}{2006}\natexlab{}.
\newblock \showarticletitle{An evaluation of similarity coefficients for
  software fault localization}. In \bibinfo{booktitle}{\emph{2006 12th Pacific
  Rim International Symposium on Dependable Computing (PRDC'06)}}. IEEE,
  \bibinfo{pages}{39--46}.
\newblock


\bibitem[\protect\citeauthoryear{An}{An}{2022}]%
        {fdg-artifact}
\bibfield{author}{\bibinfo{person}{Gabin An}.} \bibinfo{year}{2022}\natexlab{}.
\newblock \showarticletitle{agb94/FDG-artifact}.
\newblock  (\bibinfo{date}{May} \bibinfo{year}{2022}).
\newblock
\urldef\tempurl%
\url{https://doi.org/10.5281/zenodo.6525227}
\showDOI{\tempurl}


\bibitem[\protect\citeauthoryear{Artzi, Dolby, Tip, and Pistoia}{Artzi
  et~al\mbox{.}}{2010}]%
        {artzi2010directed}
\bibfield{author}{\bibinfo{person}{Shay Artzi}, \bibinfo{person}{Julian Dolby},
  \bibinfo{person}{Frank Tip}, {and} \bibinfo{person}{Marco Pistoia}.}
  \bibinfo{year}{2010}\natexlab{}.
\newblock \showarticletitle{Directed test generation for effective fault
  localization}. In \bibinfo{booktitle}{\emph{Proceedings of the 19th
  international symposium on Software testing and analysis}}.
  \bibinfo{pages}{49--60}.
\newblock


\bibitem[\protect\citeauthoryear{B.~Le, Lo, Le~Goues, and Grunske}{B.~Le
  et~al\mbox{.}}{2016}]%
        {B.-Le:2016yu}
\bibfield{author}{\bibinfo{person}{Tien-Duy B.~Le}, \bibinfo{person}{David Lo},
  \bibinfo{person}{Claire Le~Goues}, {and} \bibinfo{person}{Lars Grunske}.}
  \bibinfo{year}{2016}\natexlab{}.
\newblock \showarticletitle{A Learning-to-rank Based Fault Localization
  Approach Using Likely Invariants}. In \bibinfo{booktitle}{\emph{Proceedings
  of the 25th International Symposium on Software Testing and Analysis}}
  \emph{(\bibinfo{series}{ISSTA 2016})}. \bibinfo{publisher}{ACM},
  \bibinfo{address}{New York, NY, USA}, \bibinfo{pages}{177--188}.
\newblock


\bibitem[\protect\citeauthoryear{Bandyopadhyay and Ghosh}{Bandyopadhyay and
  Ghosh}{2011}]%
        {bandyopadhyay2011proximity}
\bibfield{author}{\bibinfo{person}{Aritra Bandyopadhyay} {and}
  \bibinfo{person}{Sudipto Ghosh}.} \bibinfo{year}{2011}\natexlab{}.
\newblock \showarticletitle{Proximity based weighting of test cases to improve
  spectrum based fault localization}. In \bibinfo{booktitle}{\emph{2011 26th
  IEEE/ACM International Conference on Automated Software Engineering (ASE
  2011)}}. IEEE, \bibinfo{pages}{420--423}.
\newblock


\bibitem[\protect\citeauthoryear{Bandyopadhyay and Ghosh}{Bandyopadhyay and
  Ghosh}{2012}]%
        {bandyopadhyay2012tester}
\bibfield{author}{\bibinfo{person}{Aritra Bandyopadhyay} {and}
  \bibinfo{person}{Sudipto Ghosh}.} \bibinfo{year}{2012}\natexlab{}.
\newblock \showarticletitle{Tester feedback driven fault localization}. In
  \bibinfo{booktitle}{\emph{2012 IEEE Fifth International Conference on
  Software Testing, Verification and Validation}}. IEEE,
  \bibinfo{pages}{41--50}.
\newblock


\bibitem[\protect\citeauthoryear{Barr, Harman, McMinn, Shahbaz, and Yoo}{Barr
  et~al\mbox{.}}{2015}]%
        {Barr:2015qd}
\bibfield{author}{\bibinfo{person}{Earl Barr}, \bibinfo{person}{Mark Harman},
  \bibinfo{person}{Phil McMinn}, \bibinfo{person}{Muzammil Shahbaz}, {and}
  \bibinfo{person}{Shin Yoo}.} \bibinfo{year}{2015}\natexlab{}.
\newblock \showarticletitle{The Oracle Problem in Software Testing: A Survey}.
\newblock \bibinfo{journal}{\emph{IEEE Transactions on Software Engineering}}
  \bibinfo{volume}{41}, \bibinfo{number}{5} (\bibinfo{date}{May}
  \bibinfo{year}{2015}), \bibinfo{pages}{507--525}.
\newblock


\bibitem[\protect\citeauthoryear{Baudry, Fleurey, and Le~Traon}{Baudry
  et~al\mbox{.}}{2006}]%
        {baudry2006improving}
\bibfield{author}{\bibinfo{person}{Benoit Baudry}, \bibinfo{person}{Franck
  Fleurey}, {and} \bibinfo{person}{Yves Le~Traon}.}
  \bibinfo{year}{2006}\natexlab{}.
\newblock \showarticletitle{Improving test suites for efficient fault
  localization}. In \bibinfo{booktitle}{\emph{Proceedings of the 28th
  international conference on Software engineering}}. \bibinfo{pages}{82--91}.
\newblock


\bibitem[\protect\citeauthoryear{B{\"o}hme, Geethal, and Pham}{B{\"o}hme
  et~al\mbox{.}}{2019}]%
        {bohme2019human}
\bibfield{author}{\bibinfo{person}{Marcel B{\"o}hme}, \bibinfo{person}{Charaka
  Geethal}, {and} \bibinfo{person}{Van-Thuan Pham}.}
  \bibinfo{year}{2019}\natexlab{}.
\newblock \showarticletitle{Human-In-The-Loop Automatic Program Repair}.
\newblock \bibinfo{journal}{\emph{arXiv preprint arXiv:1912.07758}}
  (\bibinfo{year}{2019}).
\newblock


\bibitem[\protect\citeauthoryear{Caballero, Riesco, and Silva}{Caballero
  et~al\mbox{.}}{2017}]%
        {Caballero2017rz}
\bibfield{author}{\bibinfo{person}{Rafael Caballero},
  \bibinfo{person}{Adri\'{a}n Riesco}, {and} \bibinfo{person}{Josep Silva}.}
  \bibinfo{year}{2017}\natexlab{}.
\newblock \showarticletitle{A Survey of Algorithmic Debugging}.
\newblock \bibinfo{journal}{\emph{ACM Comput. Surv.}} \bibinfo{volume}{50},
  \bibinfo{number}{4} (\bibinfo{date}{aug} \bibinfo{year}{2017}).
\newblock


\bibitem[\protect\citeauthoryear{Campos, Abreu, Fraser, and d'Amorim}{Campos
  et~al\mbox{.}}{2013}]%
        {campos2013entropy}
\bibfield{author}{\bibinfo{person}{Jos{\'e} Campos}, \bibinfo{person}{Rui
  Abreu}, \bibinfo{person}{Gordon Fraser}, {and} \bibinfo{person}{Marcelo
  d'Amorim}.} \bibinfo{year}{2013}\natexlab{}.
\newblock \showarticletitle{Entropy-based test generation for improved fault
  localization}. In \bibinfo{booktitle}{\emph{2013 28th IEEE/ACM International
  Conference on Automated Software Engineering (ASE)}}. IEEE,
  \bibinfo{pages}{257--267}.
\newblock


\bibitem[\protect\citeauthoryear{Elbaum, Malishevsky, and Rothermel}{Elbaum
  et~al\mbox{.}}{2002}]%
        {elbaum2002test}
\bibfield{author}{\bibinfo{person}{Sebastian Elbaum}, \bibinfo{person}{Alexey~G
  Malishevsky}, {and} \bibinfo{person}{Gregg Rothermel}.}
  \bibinfo{year}{2002}\natexlab{}.
\newblock \showarticletitle{Test case prioritization: A family of empirical
  studies}.
\newblock \bibinfo{journal}{\emph{IEEE transactions on software engineering}}
  \bibinfo{volume}{28}, \bibinfo{number}{2} (\bibinfo{year}{2002}),
  \bibinfo{pages}{159--182}.
\newblock


\bibitem[\protect\citeauthoryear{Fraser and Arcuri}{Fraser and Arcuri}{2011}]%
        {fraser2011evosuite}
\bibfield{author}{\bibinfo{person}{Gordon Fraser} {and} \bibinfo{person}{Andrea
  Arcuri}.} \bibinfo{year}{2011}\natexlab{}.
\newblock \showarticletitle{Evosuite: automatic test suite generation for
  object-oriented software}. In \bibinfo{booktitle}{\emph{Proceedings of the
  19th ACM SIGSOFT symposium and the 13th European conference on Foundations of
  software engineering}}. \bibinfo{pages}{416--419}.
\newblock


\bibitem[\protect\citeauthoryear{{Fraser} and {Zeller}}{{Fraser} and
  {Zeller}}{2012}]%
        {Fraser2012fr}
\bibfield{author}{\bibinfo{person}{G. {Fraser}} {and} \bibinfo{person}{A.
  {Zeller}}.} \bibinfo{year}{2012}\natexlab{}.
\newblock \showarticletitle{Mutation-Driven Generation of Unit Tests and
  Oracles}.
\newblock \bibinfo{journal}{\emph{IEEE Transactions on Software Engineering}}
  \bibinfo{volume}{38}, \bibinfo{number}{2} (\bibinfo{year}{2012}),
  \bibinfo{pages}{278--292}.
\newblock


\bibitem[\protect\citeauthoryear{Gong, Lo, Jiang, and Zhang}{Gong
  et~al\mbox{.}}{2012}]%
        {gong2012interactive}
\bibfield{author}{\bibinfo{person}{Liang Gong}, \bibinfo{person}{David Lo},
  \bibinfo{person}{Lingxiao Jiang}, {and} \bibinfo{person}{Hongyu Zhang}.}
  \bibinfo{year}{2012}\natexlab{}.
\newblock \showarticletitle{Interactive fault localization leveraging simple
  user feedback}. In \bibinfo{booktitle}{\emph{2012 28th IEEE International
  Conference on Software Maintenance (ICSM)}}. IEEE, \bibinfo{pages}{67--76}.
\newblock


\bibitem[\protect\citeauthoryear{Gonzalez-Sanchez, Abreu, Gross, and van
  Gemund}{Gonzalez-Sanchez et~al\mbox{.}}{2011a}]%
        {gonzalez2011prioritizing}
\bibfield{author}{\bibinfo{person}{Alberto Gonzalez-Sanchez},
  \bibinfo{person}{Rui Abreu}, \bibinfo{person}{Hans-Gerhard Gross}, {and}
  \bibinfo{person}{Arjan~JC van Gemund}.} \bibinfo{year}{2011}\natexlab{a}.
\newblock \showarticletitle{Prioritizing tests for fault localization through
  ambiguity group reduction}. In \bibinfo{booktitle}{\emph{2011 26th IEEE/ACM
  International Conference on Automated Software Engineering (ASE 2011)}}.
  IEEE, \bibinfo{pages}{83--92}.
\newblock


\bibitem[\protect\citeauthoryear{Gonzalez-Sanchez, Piel, Abreu, Gross, and van
  Gemund}{Gonzalez-Sanchez et~al\mbox{.}}{2011b}]%
        {Gonzalez-Sanchez:2011yq}
\bibfield{author}{\bibinfo{person}{Alberto Gonzalez-Sanchez},
  \bibinfo{person}{{\'E}ric Piel}, \bibinfo{person}{Rui Abreu},
  \bibinfo{person}{Hans-Gerhard Gross}, {and} \bibinfo{person}{Arjan J.~C. van
  Gemund}.} \bibinfo{year}{2011}\natexlab{b}.
\newblock \showarticletitle{Prioritizing tests for software fault diagnosis}.
\newblock \bibinfo{journal}{\emph{Software: Practice and Experience}}
  \bibinfo{volume}{41}, \bibinfo{number}{10} (\bibinfo{year}{2011}),
  \bibinfo{pages}{1105--1129}.
\newblock
\showISSN{1097-024X}
\urldef\tempurl%
\url{https://doi.org/10.1002/spe.1065}
\showDOI{\tempurl}


\bibitem[\protect\citeauthoryear{Gonzalez-Sanchez, Piel, Gross, and van
  Gemund}{Gonzalez-Sanchez et~al\mbox{.}}{2010}]%
        {gonzalez2010prioritizing}
\bibfield{author}{\bibinfo{person}{Alberto Gonzalez-Sanchez},
  \bibinfo{person}{Eric Piel}, \bibinfo{person}{Hans-Gerhard Gross}, {and}
  \bibinfo{person}{Arjan~JC van Gemund}.} \bibinfo{year}{2010}\natexlab{}.
\newblock \showarticletitle{Prioritizing tests for software fault
  localization}. In \bibinfo{booktitle}{\emph{2010 10th International
  Conference on Quality Software}}. IEEE, \bibinfo{pages}{42--51}.
\newblock


\bibitem[\protect\citeauthoryear{Halmos}{Halmos}{2017}]%
        {halmos2017naive}
\bibfield{author}{\bibinfo{person}{Paul~R Halmos}.}
  \bibinfo{year}{2017}\natexlab{}.
\newblock \bibinfo{booktitle}{\emph{Naive set theory}}.
\newblock \bibinfo{publisher}{Courier Dover Publications}.
\newblock


\bibitem[\protect\citeauthoryear{Hao, Xie, Zhang, Wang, Sun, and Mei}{Hao
  et~al\mbox{.}}{2010}]%
        {hao2010test}
\bibfield{author}{\bibinfo{person}{Dan Hao}, \bibinfo{person}{Tao Xie},
  \bibinfo{person}{Lu Zhang}, \bibinfo{person}{Xiaoyin Wang},
  \bibinfo{person}{Jiasu Sun}, {and} \bibinfo{person}{Hong Mei}.}
  \bibinfo{year}{2010}\natexlab{}.
\newblock \showarticletitle{Test input reduction for result inspection to
  facilitate fault localization}.
\newblock \bibinfo{journal}{\emph{Automated software engineering}}
  \bibinfo{volume}{17}, \bibinfo{number}{1} (\bibinfo{year}{2010}),
  \bibinfo{pages}{5}.
\newblock


\bibitem[\protect\citeauthoryear{Hao, Zhang, Xie, Mei, and Sun}{Hao
  et~al\mbox{.}}{2009a}]%
        {hao2009interactive}
\bibfield{author}{\bibinfo{person}{Dan Hao}, \bibinfo{person}{Lu Zhang},
  \bibinfo{person}{Tao Xie}, \bibinfo{person}{Hong Mei}, {and}
  \bibinfo{person}{Jia-Su Sun}.} \bibinfo{year}{2009}\natexlab{a}.
\newblock \showarticletitle{Interactive fault localization using test
  information}.
\newblock \bibinfo{journal}{\emph{Journal of Computer Science and Technology}}
  \bibinfo{volume}{24}, \bibinfo{number}{5} (\bibinfo{year}{2009}),
  \bibinfo{pages}{962--974}.
\newblock


\bibitem[\protect\citeauthoryear{Hao, Zhang, Zhang, Sun, and Mei}{Hao
  et~al\mbox{.}}{2009b}]%
        {hao2009vida}
\bibfield{author}{\bibinfo{person}{Dan Hao}, \bibinfo{person}{Lingming Zhang},
  \bibinfo{person}{Lu Zhang}, \bibinfo{person}{Jiasu Sun}, {and}
  \bibinfo{person}{Hong Mei}.} \bibinfo{year}{2009}\natexlab{b}.
\newblock \showarticletitle{VIDA: Visual interactive debugging}. In
  \bibinfo{booktitle}{\emph{2009 IEEE 31st International Conference on Software
  Engineering}}. IEEE, \bibinfo{pages}{583--586}.
\newblock


\bibitem[\protect\citeauthoryear{Hong, Kwak, Lee, Jeon, Ko, Kim, and Kim}{Hong
  et~al\mbox{.}}{2017}]%
        {Hong:2017qy}
\bibfield{author}{\bibinfo{person}{Shin Hong}, \bibinfo{person}{Taehoon Kwak},
  \bibinfo{person}{Byeongcheol Lee}, \bibinfo{person}{Yiru Jeon},
  \bibinfo{person}{Bongsuk Ko}, \bibinfo{person}{Yunho Kim}, {and}
  \bibinfo{person}{Moonzoo Kim}.} \bibinfo{year}{2017}\natexlab{}.
\newblock \showarticletitle{MUSEUM: Debugging Real-World Multilingual Programs
  Using Mutation Analysis}.
\newblock \bibinfo{journal}{\emph{Information and Software Technology}}
  \bibinfo{volume}{82} (\bibinfo{year}{2017}), \bibinfo{pages}{80--95}.
\newblock


\bibitem[\protect\citeauthoryear{Jin and Orso}{Jin and Orso}{2012}]%
        {jin2012bugredux}
\bibfield{author}{\bibinfo{person}{Wei Jin} {and} \bibinfo{person}{Alessandro
  Orso}.} \bibinfo{year}{2012}\natexlab{}.
\newblock \showarticletitle{BugRedux: reproducing field failures for in-house
  debugging}. In \bibinfo{booktitle}{\emph{2012 34th International Conference
  on Software Engineering (ICSE)}}. IEEE, \bibinfo{pages}{474--484}.
\newblock


\bibitem[\protect\citeauthoryear{Jin and Orso}{Jin and Orso}{2013}]%
        {jin2013f3}
\bibfield{author}{\bibinfo{person}{Wei Jin} {and} \bibinfo{person}{Alessandro
  Orso}.} \bibinfo{year}{2013}\natexlab{}.
\newblock \showarticletitle{F3: fault localization for field failures}. In
  \bibinfo{booktitle}{\emph{Proceedings of the 2013 International Symposium on
  Software Testing and Analysis}}. \bibinfo{pages}{213--223}.
\newblock


\bibitem[\protect\citeauthoryear{Jones and Harrold}{Jones and Harrold}{2005}]%
        {jones2005empirical}
\bibfield{author}{\bibinfo{person}{James~A Jones} {and}
  \bibinfo{person}{Mary~Jean Harrold}.} \bibinfo{year}{2005}\natexlab{}.
\newblock \showarticletitle{Empirical evaluation of the tarantula automatic
  fault-localization technique}. In \bibinfo{booktitle}{\emph{Proceedings of
  the 20th IEEE/ACM international Conference on Automated software
  engineering}}. \bibinfo{pages}{273--282}.
\newblock


\bibitem[\protect\citeauthoryear{Jost}{Jost}{2006}]%
        {jost2006entropy}
\bibfield{author}{\bibinfo{person}{Lou Jost}.} \bibinfo{year}{2006}\natexlab{}.
\newblock \showarticletitle{Entropy and diversity}.
\newblock \bibinfo{journal}{\emph{Oikos}} \bibinfo{volume}{113},
  \bibinfo{number}{2} (\bibinfo{year}{2006}), \bibinfo{pages}{363--375}.
\newblock


\bibitem[\protect\citeauthoryear{Just, Jalali, and Ernst}{Just
  et~al\mbox{.}}{2014}]%
        {just2014defects4j}
\bibfield{author}{\bibinfo{person}{Ren{\'e} Just}, \bibinfo{person}{Darioush
  Jalali}, {and} \bibinfo{person}{Michael~D Ernst}.}
  \bibinfo{year}{2014}\natexlab{}.
\newblock \showarticletitle{Defects4J: A database of existing faults to enable
  controlled testing studies for Java programs}. In
  \bibinfo{booktitle}{\emph{Proceedings of the 2014 International Symposium on
  Software Testing and Analysis}}. \bibinfo{pages}{437--440}.
\newblock


\bibitem[\protect\citeauthoryear{Ko and Myers}{Ko and Myers}{2008}]%
        {ko2008debugging}
\bibfield{author}{\bibinfo{person}{Andrew Ko} {and} \bibinfo{person}{Brad
  Myers}.} \bibinfo{year}{2008}\natexlab{}.
\newblock \showarticletitle{Debugging reinvented}. In
  \bibinfo{booktitle}{\emph{2008 ACM/IEEE 30th International Conference on
  Software Engineering}}. IEEE, \bibinfo{pages}{301--310}.
\newblock


\bibitem[\protect\citeauthoryear{Ko and Myers}{Ko and Myers}{2004}]%
        {ko2004designing}
\bibfield{author}{\bibinfo{person}{Andrew~J Ko} {and} \bibinfo{person}{Brad~A
  Myers}.} \bibinfo{year}{2004}\natexlab{}.
\newblock \showarticletitle{Designing the whyline: a debugging interface for
  asking questions about program behavior}. In
  \bibinfo{booktitle}{\emph{Proceedings of the SIGCHI conference on Human
  factors in computing systems}}. \bibinfo{pages}{151--158}.
\newblock


\bibitem[\protect\citeauthoryear{Kuma, Higo, Matsumoto, and Kusumoto}{Kuma
  et~al\mbox{.}}{2020}]%
        {kuma2020improving}
\bibfield{author}{\bibinfo{person}{Tetsushi Kuma}, \bibinfo{person}{Yoshiki
  Higo}, \bibinfo{person}{Shinsuke Matsumoto}, {and} \bibinfo{person}{Shinji
  Kusumoto}.} \bibinfo{year}{2020}\natexlab{}.
\newblock \showarticletitle{Improving the Accuracy of Spectrum-based Fault
  Localization for Automated Program Repair}. In
  \bibinfo{booktitle}{\emph{Proceedings of the 28th International Conference on
  Program Comprehension}}. \bibinfo{pages}{376--380}.
\newblock


\bibitem[\protect\citeauthoryear{Li, d'Amorim, and Orso}{Li
  et~al\mbox{.}}{2016}]%
        {li2016iterative}
\bibfield{author}{\bibinfo{person}{Xiangyu Li}, \bibinfo{person}{Marcelo
  d'Amorim}, {and} \bibinfo{person}{Alessandro Orso}.}
  \bibinfo{year}{2016}\natexlab{}.
\newblock \showarticletitle{Iterative user-driven fault localization}. In
  \bibinfo{booktitle}{\emph{Haifa Verification Conference}}. Springer,
  \bibinfo{pages}{82--98}.
\newblock


\bibitem[\protect\citeauthoryear{Li, Li, Zhang, and Zhang}{Li
  et~al\mbox{.}}{2019}]%
        {Li2019aa}
\bibfield{author}{\bibinfo{person}{Xia Li}, \bibinfo{person}{Wei Li},
  \bibinfo{person}{Yuqun Zhang}, {and} \bibinfo{person}{Lingming Zhang}.}
  \bibinfo{year}{2019}\natexlab{}.
\newblock \showarticletitle{DeepFL: Integrating Multiple Fault Diagnosis
  Dimensions for Deep Fault Localization}. In
  \bibinfo{booktitle}{\emph{Proceedings of the 28th ACM SIGSOFT International
  Symposium on Software Testing and Analysis}} \emph{(\bibinfo{series}{ISSTA
  2019})}. \bibinfo{publisher}{Association for Computing Machinery},
  \bibinfo{address}{New York, NY, USA}, \bibinfo{pages}{169--180}.
\newblock


\bibitem[\protect\citeauthoryear{Li, Zhu, d'Amorim, and Orso}{Li
  et~al\mbox{.}}{2018}]%
        {li2018enlightened}
\bibfield{author}{\bibinfo{person}{Xiangyu Li}, \bibinfo{person}{Shaowei Zhu},
  \bibinfo{person}{Marcelo d'Amorim}, {and} \bibinfo{person}{Alessandro Orso}.}
  \bibinfo{year}{2018}\natexlab{}.
\newblock \showarticletitle{Enlightened debugging}. In
  \bibinfo{booktitle}{\emph{2018 IEEE/ACM 40th International Conference on
  Software Engineering (ICSE)}}. IEEE, \bibinfo{pages}{82--92}.
\newblock


\bibitem[\protect\citeauthoryear{Lin, Sun, Xue, Liu, and Dong}{Lin
  et~al\mbox{.}}{2017}]%
        {lin2017feedback}
\bibfield{author}{\bibinfo{person}{Yun Lin}, \bibinfo{person}{Jun Sun},
  \bibinfo{person}{Yinxing Xue}, \bibinfo{person}{Yang Liu}, {and}
  \bibinfo{person}{Jinsong Dong}.} \bibinfo{year}{2017}\natexlab{}.
\newblock \showarticletitle{Feedback-based debugging}. In
  \bibinfo{booktitle}{\emph{2017 IEEE/ACM 39th International Conference on
  Software Engineering (ICSE)}}. IEEE, \bibinfo{pages}{393--403}.
\newblock


\bibitem[\protect\citeauthoryear{Masri, Abou-Assi, El-Ghali, and
  Al-Fatairi}{Masri et~al\mbox{.}}{2009}]%
        {masri2009empirical}
\bibfield{author}{\bibinfo{person}{Wes Masri}, \bibinfo{person}{Rawad
  Abou-Assi}, \bibinfo{person}{Marwa El-Ghali}, {and} \bibinfo{person}{Nour
  Al-Fatairi}.} \bibinfo{year}{2009}\natexlab{}.
\newblock \showarticletitle{An empirical study of the factors that reduce the
  effectiveness of coverage-based fault localization}. In
  \bibinfo{booktitle}{\emph{Proceedings of the 2nd International Workshop on
  Defects in Large Software Systems: Held in conjunction with the ACM SIGSOFT
  International Symposium on Software Testing and Analysis (ISSTA 2009)}}.
  \bibinfo{pages}{1--5}.
\newblock


\bibitem[\protect\citeauthoryear{Moon, Kim, Kim, and Yoo}{Moon
  et~al\mbox{.}}{2014}]%
        {Moon:2014ly}
\bibfield{author}{\bibinfo{person}{Seokhyeon Moon}, \bibinfo{person}{Yunho
  Kim}, \bibinfo{person}{Moonzoo Kim}, {and} \bibinfo{person}{Shin Yoo}.}
  \bibinfo{year}{2014}\natexlab{}.
\newblock \showarticletitle{Ask the Mutants: Mutating Faulty Programs for Fault
  Localization}. In \bibinfo{booktitle}{\emph{Proceedings of the 7th
  International Conference on Software Testing, Verification and Validation}}
  \emph{(\bibinfo{series}{ICST 2014})}. \bibinfo{pages}{153--162}.
\newblock


\bibitem[\protect\citeauthoryear{Naish, Lee, and Ramamohanarao}{Naish
  et~al\mbox{.}}{2011}]%
        {naish2011model}
\bibfield{author}{\bibinfo{person}{Lee Naish}, \bibinfo{person}{Hua~Jie Lee},
  {and} \bibinfo{person}{Kotagiri Ramamohanarao}.}
  \bibinfo{year}{2011}\natexlab{}.
\newblock \showarticletitle{A model for spectra-based software diagnosis}.
\newblock \bibinfo{journal}{\emph{ACM Transactions on software engineering and
  methodology (TOSEM)}} \bibinfo{volume}{20}, \bibinfo{number}{3}
  (\bibinfo{year}{2011}), \bibinfo{pages}{1--32}.
\newblock


\bibitem[\protect\citeauthoryear{Pacheco and Ernst}{Pacheco and Ernst}{2007}]%
        {pacheco2007randoop}
\bibfield{author}{\bibinfo{person}{Carlos Pacheco} {and}
  \bibinfo{person}{Michael~D Ernst}.} \bibinfo{year}{2007}\natexlab{}.
\newblock \showarticletitle{Randoop: feedback-directed random testing for
  Java}. In \bibinfo{booktitle}{\emph{Companion to the 22nd ACM SIGPLAN
  conference on Object-oriented programming systems and applications
  companion}}. \bibinfo{pages}{815--816}.
\newblock


\bibitem[\protect\citeauthoryear{Papadakis and Le-Traon}{Papadakis and
  Le-Traon}{2012}]%
        {Papadakis:2012fk}
\bibfield{author}{\bibinfo{person}{M. Papadakis} {and} \bibinfo{person}{Y.
  Le-Traon}.} \bibinfo{year}{2012}\natexlab{}.
\newblock \showarticletitle{Using Mutants to Locate "Unknown" Faults}. In
  \bibinfo{booktitle}{\emph{Proceedings of the 5th IEEE Fifth International
  Conference on Software Testing, Verification and Validation}}
  \emph{(\bibinfo{series}{Mutation 2012})}. \bibinfo{pages}{691--700}.
\newblock
\urldef\tempurl%
\url{https://doi.org/10.1109/ICST.2012.159}
\showDOI{\tempurl}


\bibitem[\protect\citeauthoryear{Papadakis and Traon}{Papadakis and
  Traon}{2015}]%
        {Papadakis:2015sf}
\bibfield{author}{\bibinfo{person}{Mike Papadakis} {and}
  \bibinfo{person}{Yves~Le Traon}.} \bibinfo{year}{2015}\natexlab{}.
\newblock \showarticletitle{Metallaxis-FL: mutation-based fault localization}.
\newblock \bibinfo{journal}{\emph{Softw. Test., Verif. Reliab.}}
  \bibinfo{volume}{25}, \bibinfo{number}{5-7} (\bibinfo{year}{2015}),
  \bibinfo{pages}{605--628}.
\newblock
\urldef\tempurl%
\url{https://doi.org/10.1002/stvr.1509}
\showDOI{\tempurl}


\bibitem[\protect\citeauthoryear{Parnin and Orso}{Parnin and Orso}{2011}]%
        {parnin2011automated}
\bibfield{author}{\bibinfo{person}{Chris Parnin} {and}
  \bibinfo{person}{Alessandro Orso}.} \bibinfo{year}{2011}\natexlab{}.
\newblock \showarticletitle{Are automated debugging techniques actually helping
  programmers?}. In \bibinfo{booktitle}{\emph{Proceedings of the 2011
  international symposium on software testing and analysis}}.
  \bibinfo{pages}{199--209}.
\newblock


\bibitem[\protect\citeauthoryear{Pastore, Mariani, and Fraser}{Pastore
  et~al\mbox{.}}{2013}]%
        {pastore2013crowdoracles}
\bibfield{author}{\bibinfo{person}{Fabrizio Pastore}, \bibinfo{person}{Leonardo
  Mariani}, {and} \bibinfo{person}{Gordon Fraser}.}
  \bibinfo{year}{2013}\natexlab{}.
\newblock \showarticletitle{Crowdoracles: Can the crowd solve the oracle
  problem?}. In \bibinfo{booktitle}{\emph{2013 IEEE Sixth International
  Conference on Software Testing, Verification and Validation}}. IEEE,
  \bibinfo{pages}{342--351}.
\newblock


\bibitem[\protect\citeauthoryear{Perez, Abreu, and van Deursen}{Perez
  et~al\mbox{.}}{2017}]%
        {perez2017test}
\bibfield{author}{\bibinfo{person}{Alexandre Perez}, \bibinfo{person}{Rui
  Abreu}, {and} \bibinfo{person}{Arie van Deursen}.}
  \bibinfo{year}{2017}\natexlab{}.
\newblock \showarticletitle{A test-suite diagnosability metric for
  spectrum-based fault localization approaches}. In
  \bibinfo{booktitle}{\emph{2017 IEEE/ACM 39th International Conference on
  Software Engineering (ICSE)}}. IEEE, \bibinfo{pages}{654--664}.
\newblock


\bibitem[\protect\citeauthoryear{Renieres and Reiss}{Renieres and
  Reiss}{2003}]%
        {renieres2003fault}
\bibfield{author}{\bibinfo{person}{Manos Renieres} {and}
  \bibinfo{person}{Steven~P Reiss}.} \bibinfo{year}{2003}\natexlab{}.
\newblock \showarticletitle{Fault localization with nearest neighbor queries}.
  In \bibinfo{booktitle}{\emph{18th IEEE International Conference on Automated
  Software Engineering, 2003. Proceedings.}} IEEE, \bibinfo{pages}{30--39}.
\newblock


\bibitem[\protect\citeauthoryear{R{\"o}$\beta$ler, Fraser, Zeller, and
  Orso}{R{\"o}$\beta$ler et~al\mbox{.}}{2012}]%
        {robetaler2012isolating}
\bibfield{author}{\bibinfo{person}{Jeremias R{\"o}$\beta$ler},
  \bibinfo{person}{Gordon Fraser}, \bibinfo{person}{Andreas Zeller}, {and}
  \bibinfo{person}{Alessandro Orso}.} \bibinfo{year}{2012}\natexlab{}.
\newblock \showarticletitle{Isolating failure causes through test case
  generation}. In \bibinfo{booktitle}{\emph{Proceedings of the 2012
  international symposium on software testing and analysis}}.
  \bibinfo{pages}{309--319}.
\newblock


\bibitem[\protect\citeauthoryear{Rothermel, Untch, Chu, and Harrold}{Rothermel
  et~al\mbox{.}}{1999}]%
        {rothermel1999test}
\bibfield{author}{\bibinfo{person}{Gregg Rothermel}, \bibinfo{person}{Roland~H
  Untch}, \bibinfo{person}{Chengyun Chu}, {and} \bibinfo{person}{Mary~Jean
  Harrold}.} \bibinfo{year}{1999}\natexlab{}.
\newblock \showarticletitle{Test case prioritization: An empirical study}. In
  \bibinfo{booktitle}{\emph{Proceedings IEEE International Conference on
  Software Maintenance-1999 (ICSM'99).'Software Maintenance for Business
  Change'(Cat. No. 99CB36360)}}. IEEE, \bibinfo{pages}{179--188}.
\newblock


\bibitem[\protect\citeauthoryear{Saha, Lease, Khurshid, and Perry}{Saha
  et~al\mbox{.}}{2013}]%
        {Saha:2013rm}
\bibfield{author}{\bibinfo{person}{Ripon~K Saha}, \bibinfo{person}{Matthew
  Lease}, \bibinfo{person}{Sarfraz Khurshid}, {and} \bibinfo{person}{Dewayne~E
  Perry}.} \bibinfo{year}{2013}\natexlab{}.
\newblock \showarticletitle{Improving bug localization using structured
  information retrieval}. In \bibinfo{booktitle}{\emph{Automated Software
  Engineering (ASE), 2013 IEEE/ACM 28th International Conference on}}. IEEE,
  \bibinfo{pages}{345--355}.
\newblock


\bibitem[\protect\citeauthoryear{Shannon}{Shannon}{1948}]%
        {shannon1948mathematical}
\bibfield{author}{\bibinfo{person}{Claude~E Shannon}.}
  \bibinfo{year}{1948}\natexlab{}.
\newblock \showarticletitle{A mathematical theory of communication}.
\newblock \bibinfo{journal}{\emph{The Bell system technical journal}}
  \bibinfo{volume}{27}, \bibinfo{number}{3} (\bibinfo{year}{1948}),
  \bibinfo{pages}{379--423}.
\newblock


\bibitem[\protect\citeauthoryear{Shapiro}{Shapiro}{1982}]%
        {Shapiro1982jk}
\bibfield{author}{\bibinfo{person}{Ehud~Y. Shapiro}.}
  \bibinfo{year}{1982}\natexlab{}.
\newblock \showarticletitle{Algorithmic Program Diagnosis}. In
  \bibinfo{booktitle}{\emph{Proceedings of the 9th ACM SIGPLAN-SIGACT Symposium
  on Principles of Programming Languages}} \emph{(\bibinfo{series}{POPL '82})}.
  \bibinfo{publisher}{Association for Computing Machinery},
  \bibinfo{address}{New York, NY, USA}, \bibinfo{pages}{299--308}.
\newblock


\bibitem[\protect\citeauthoryear{Sohn and Yoo}{Sohn and Yoo}{2017}]%
        {Sohn:2017xq}
\bibfield{author}{\bibinfo{person}{Jeongju Sohn} {and} \bibinfo{person}{Shin
  Yoo}.} \bibinfo{year}{2017}\natexlab{}.
\newblock \showarticletitle{FLUCCS: Using Code and Change Metrics to Improve
  Fault Localisation}. In \bibinfo{booktitle}{\emph{Proceedings of the
  International Symposium on Software Testing and Analysis}}
  \emph{(\bibinfo{series}{ISSTA 2017})}. \bibinfo{pages}{273--283}.
\newblock


\bibitem[\protect\citeauthoryear{Stenbakken, Souders, and Stewart}{Stenbakken
  et~al\mbox{.}}{1989}]%
        {stenbakken1989ambiguity}
\bibfield{author}{\bibinfo{person}{GN Stenbakken}, \bibinfo{person}{TM
  Souders}, {and} \bibinfo{person}{GW Stewart}.}
  \bibinfo{year}{1989}\natexlab{}.
\newblock \showarticletitle{Ambiguity groups and testability}.
\newblock \bibinfo{journal}{\emph{IEEE Transactions on Instrumentation and
  Measurement}} \bibinfo{volume}{38}, \bibinfo{number}{5}
  (\bibinfo{year}{1989}), \bibinfo{pages}{941--947}.
\newblock


\bibitem[\protect\citeauthoryear{{Wong}, {Debroy}, {Gao}, and {Li}}{{Wong}
  et~al\mbox{.}}{2014}]%
        {wong2014tor}
\bibfield{author}{\bibinfo{person}{W.~E. {Wong}}, \bibinfo{person}{V.
  {Debroy}}, \bibinfo{person}{R. {Gao}}, {and} \bibinfo{person}{Y. {Li}}.}
  \bibinfo{year}{2014}\natexlab{}.
\newblock \showarticletitle{The DStar Method for Effective Software Fault
  Localization}.
\newblock \bibinfo{journal}{\emph{IEEE Transactions on Reliability}}
  \bibinfo{volume}{63}, \bibinfo{number}{1} (\bibinfo{year}{2014}),
  \bibinfo{pages}{290--308}.
\newblock


\bibitem[\protect\citeauthoryear{Wong, Gao, Li, Abreu, and Wotawa}{Wong
  et~al\mbox{.}}{2016}]%
        {wong2016survey}
\bibfield{author}{\bibinfo{person}{W~Eric Wong}, \bibinfo{person}{Ruizhi Gao},
  \bibinfo{person}{Yihao Li}, \bibinfo{person}{Rui Abreu}, {and}
  \bibinfo{person}{Franz Wotawa}.} \bibinfo{year}{2016}\natexlab{}.
\newblock \showarticletitle{A survey on software fault localization}.
\newblock \bibinfo{journal}{\emph{IEEE Transactions on Software Engineering}}
  \bibinfo{volume}{42}, \bibinfo{number}{8} (\bibinfo{year}{2016}),
  \bibinfo{pages}{707--740}.
\newblock


\bibitem[\protect\citeauthoryear{Xia, Bao, Lo, and Li}{Xia
  et~al\mbox{.}}{2016}]%
        {Xia2016fi}
\bibfield{author}{\bibinfo{person}{X. Xia}, \bibinfo{person}{L. Bao},
  \bibinfo{person}{D. Lo}, {and} \bibinfo{person}{S. Li}.}
  \bibinfo{year}{2016}\natexlab{}.
\newblock \showarticletitle{``{A}utomated Debugging Considered Harmful''
  Considered Harmful: A User Study Revisiting the Usefulness of Spectra-Based
  Fault Localization Techniques with Professionals Using Real Bugs from Large
  Systems}. In \bibinfo{booktitle}{\emph{Proceedings of the IEEE International
  Conference on Software Maintenance and Evolution}}
  \emph{(\bibinfo{series}{ICSME 2016})}. \bibinfo{pages}{267--278}.
\newblock


\bibitem[\protect\citeauthoryear{Xuan and Monperrus}{Xuan and
  Monperrus}{2014}]%
        {xuan2014test}
\bibfield{author}{\bibinfo{person}{Jifeng Xuan} {and} \bibinfo{person}{Martin
  Monperrus}.} \bibinfo{year}{2014}\natexlab{}.
\newblock \showarticletitle{Test case purification for improving fault
  localization}. In \bibinfo{booktitle}{\emph{Proceedings of the 22nd ACM
  SIGSOFT International Symposium on Foundations of Software Engineering}}.
  \bibinfo{pages}{52--63}.
\newblock


\bibitem[\protect\citeauthoryear{Yoo and Harman}{Yoo and Harman}{2012}]%
        {yoo2012regression}
\bibfield{author}{\bibinfo{person}{Shin Yoo} {and} \bibinfo{person}{Mark
  Harman}.} \bibinfo{year}{2012}\natexlab{}.
\newblock \showarticletitle{Regression testing minimization, selection and
  prioritization: a survey}.
\newblock \bibinfo{journal}{\emph{Software testing, verification and
  reliability}} \bibinfo{volume}{22}, \bibinfo{number}{2}
  (\bibinfo{year}{2012}), \bibinfo{pages}{67--120}.
\newblock


\bibitem[\protect\citeauthoryear{Yoo, Harman, and Clark}{Yoo
  et~al\mbox{.}}{2013}]%
        {yoo2013fault}
\bibfield{author}{\bibinfo{person}{Shin Yoo}, \bibinfo{person}{Mark Harman},
  {and} \bibinfo{person}{David Clark}.} \bibinfo{year}{2013}\natexlab{}.
\newblock \showarticletitle{Fault localization prioritization: Comparing
  information-theoretic and coverage-based approaches}.
\newblock \bibinfo{journal}{\emph{ACM Transactions on Software Engineering and
  Methodology (TOSEM)}} \bibinfo{volume}{22}, \bibinfo{number}{3}
  (\bibinfo{year}{2013}), \bibinfo{pages}{1--29}.
\newblock


\end{thebibliography}

\end{document}